\documentclass{general}
\usepackage{enumerate}
\usepackage{amssymb}
\usepackage{tabularx}
\usepackage{booktabs}
\newcolumntype{Y}{>{\centering\arraybackslash}X}

\begin{document}
\setstretch{1}
\title{Joint Quantile Regression for Spatial Data}

\author{Xu Chen}
\author{Surya T. Tokdar}
\affil{Department of Statistical Science, Duke University, Durham, NC}
\date{\today}

\def\BSRE{\textsc{BSRE}}
\def\KB{\textsc{KB}}
\def\JSQR{\textsc{JSQR}}
\def\JQR{\textsc{JQR}}
\def\ALDQR{\textsc{ALDQR}}
\def\ALPQR{\textsc{ALPQR}}
\def\ASQR{\textsc{ASQR}}
\def\JQRBS{\textsc{JQR-BS}}
\def\yt{YT17}
\def\pmc{$\textrm{PM}_{2.5}$}

\maketitle

\begin{abstract}
Linear quantile regression is a powerful tool to investigate how predictors may affect a response heterogeneously across different quantile levels. Unfortunately, existing approaches find it extremely difficult to adjust for any dependency between observation units, largely because such methods are not based upon a fully generative model of the data. For analyzing spatially indexed data, we address this difficulty by generalizing the joint quantile regression model of \cite{yang2017joint} and characterizing spatial dependence via a Gaussian or $t$ copula process on the underlying quantile levels of the observation units. A Bayesian semiparametric approach is introduced to perform inference of model parameters and carry out spatial quantile smoothing. An effective model comparison criteria is provided, particularly for selecting between different model specifications of tail heaviness and tail dependence. Extensive simulation studies and an application to particulate matter concentration in northeast US are presented to illustrate substantial gains in inference quality, accuracy and uncertainty quantification over existing alternatives. 
\end{abstract}


\section{Introduction}
\label{sec:intro}
%\subsection{Quantile regression for spatial modeling}
%\label{subsec:jointqr}
In spite of the surging popularity of quantile regression \citep[QR]{koenker1978regression} and its successful application to many fields of science \citep[see][for a review]{koenker2005quantile}, scant work exists in the domain of spatial data analysis.  When observation units are spatially indexed, ignoring the possible spatial dependence of the records may lead to biased estimates of model parameters. At the same time, incorporating spatial dependence within the quantile regression framework presents a formidable challenge. 

The standard linear quantile regression ``model'' formalizes the relationship between a scalar response $Y$ and a predictor vector $X \in\mathcal{X}\subset \mathbb{R}^p$ as
\begin{align}
Q_Y(\tau\mid X)=\beta_0(\tau) +X^\top\beta(\tau),
\label{equ:lqrmodel}
\end{align}
where $Q_Y(\tau \mid X)$ denotes the conditional response quantile at a given quantile level $\tau \in (0,1)$. The unknown and level-specific intercept and slope parameters $\beta_0(\tau)$ and $\beta(\tau) = (\beta_1(\tau), \ldots, \beta_p(\tau))^\top$ may be estimated by minimizing an averaged, level-specific check loss function $\rho_\tau(\varepsilon) = \varepsilon\{\tau - \mathds{1}(\varepsilon < 0)\}$. The resulting analysis is more robust against heteroskedasticity and outliers than least squares regression. But more importantly, through the choice of an appropriate quantile level $\tau$, the analysis can also reveal predictor effects that may be present only in the tails, thus capturing a richer variety of dependence than what is possible under ordinary mean or median regression. 

Unfortunately, the level-specific formulation \eqref{equ:lqrmodel} does not lead to any obvious probabilistic, generative model for an individual, and hence, multiple realizations of $Y$. Without the existence of such a joint generative model for the observation units, there is a considerable barrier to extending the quantile regression analysis to the spatial setting where spatial dependency must be modeled, estimated and adjusted for.

%where $\beta_0(\tau)\in\mathbb{R}$ and $\beta(\tau)\in\mathbb{R}^p$ are unknown function valued intercept and slope at $\tau$th quantile of conditional distribution of response $Y\in\mathbb{R}$ given predictor $X\in\mathcal{X}\subset\mathbb{R}^p$. Compromising between simple mean regression and full density regression, QR enriches the flexibility of the former while retains efficient computation and straightforward interpretation of predictor effect compared to the latter. Such interpretability motivates the use of QR in many scientific investigations when understanding how predictors affect the extremes of a response distribution and/or how predictor effects vary across response quantile levels are crucial \citep{chamaille2012case, burgette2011exploratory, elsner2008increasing, cade2003gentle}. Fulfilling the promise of QR requires inference over multiple quantile levels, which poses the monotonicity constraint to quantile planes, that is,
%\begin{align}
%\beta_0(\tau_1)+x^\top\beta(\tau_1)\geqslant\beta_0(\tau_2)+x^\top\beta(\tau_2)
%\label{equ:monotone}
%\end{align} 
%for any $\tau_1>\tau_2$ and any $x\in\mathcal{X}$. Ignoring this constraint and making separate inferences at many individual quantile levels result in quantile crossing and a poor information borrowing across quantiles. 

%Although there is a vast literature on QR, scant work exists on development of QR model for spatial data due to the difficulty of incorporating spatial dependence. 

\cite{hallin2009local} address this challenge with a two-stage approach where both response and predictor data are first {\it spatially detrended} via smoothing and then subjected to a (locally) linear quantile regression analysis. Their approach depends fundamentally on the assumption that spatial dependency in the observed data arises because of unobserved, spatially smooth shifts added to realizations of $X$ and $Y$. This is akin to assuming a spatial random intercept model, which is unfit to capture a rich class of plausible spatial dependency structures. 
%This method requires grid-valued sites of observations and spatial interpolation is not able to be made. 
A more comprehensive approach is offered by \cite{lum2012spatial} who embed \eqref{equ:lqrmodel} within the widely used asymmetric Laplace error regression model \citep{yu2001bayesian} and then extend it to a novel spatial process model. Notwithstanding its popularity, the asymmetric Laplace error model is known to be sensitive to heteroskedasticity and outliers, and may severely underestimate uncertainty in parameter estimation \citep{tokdar2012simultaneous}. Moreover, the asymmetric Laplace process model based analysis is sensitive to prior choice and the resulting statistical performance appears sub-par at quantile levels away from the median (Sections \ref{sec:experiment} and \ref{sec:case}).
%Although both preceding methods allow spatial adjustment of quantile function, the regression coefficients are not spatially varying (after detrending for \cite{hallin2009local}). These two methods are mainly designed for modeling data with potentially a large number of sites but few observations at each site. 

We consider an alternative track to address spatial dependency in quantile regression by extending the joint quantile regression framework of \cite{yang2017joint} (\yt, hereafter), who consider estimating \eqref{equ:lqrmodel} simultaneously for all $\tau \in (0,1)$. Joint estimation has a strong practical appeal since most scientific applications of quantile regression crucially depend on comparing and contrasting the slope estimates across various quantile levels \citep{cade1999estimating, machado2005counterfactual, elsner2008increasing}. The common practice of combining several separate quantile level analyses into a single report suffers from the problems of quantile crossing, and a poor borrowing of information across levels resulting in widely fluctuating standard error estimates and $p$-values \citep{tokdar2012simultaneous}. By proposing to estimate \eqref{equ:lqrmodel} jointly for all $\tau \in (0,1)$, one also gains a great theoretical advantage by automatically constructing a complete generative model for the response in the form
\begin{equation}
    Y = \beta_0(U) + X^\top\beta(U),\quad U \sim \Un(0,1),
    \label{equ:generative}
\end{equation}
as long as the model parameters satisfy the non-crossing condition:
\begin{equation}
\dot\beta_0(\tau) + x^\top\dot \beta(\tau) > 0~~\text{for all}~\tau \in (0,1)~ \text{and all}~ x \in \mathcal{X},
\label{equ:monotone}
\end{equation}
where $\dot{g}(\cdot)$ denotes the derivative of a function $g(\cdot)$.

In order to extend the model \eqref{equ:generative} to the case of spatially dependent units, we propose to model the joint distribution of the realizations of the {\it random quantile level} $U$ by a spatial copula process (Section \ref{sec:jsqr}). We consider a Gaussian copula with unknown parameters induced by a Gaussian spatial process widely used in the literature \citep{cressie1993statistics, banerjee2008gaussian, banerjee2014hierarchical} and show that our model includes the {\it basic spatial random effects model} \citep[\eqref{equ:bsre}]{cressie1993statistics} as a special case. A Bayesian semiparametric method is introduced (Section \ref{sec:computation}) to carry out parameter estimation by building upon the reparametrization trick and the adaptive Markov chain Monte Carlo (MCMC) inference algorithm of \yt. A software implementation of our method is available on \url{https://github.com/xuchenstat/JSQR}.
%When analyzing datasets with a large number of spatial locations, reduced rank approximation can be carried out using Gaussian predictive process \citep{tokdar2007towards, banerjee2008gaussian}. 
%To the best of our knowledge, the proposed method is the first fully model based approach for spatial quantile regression which allows making joint inference at multiple quantile levels with monotonicity constraint under the most general conditions on the predictor space $\mathcal{X}$.  

Extensive simulation studies are presented (Section \ref{sec:experiment}) to demonstrate superior statistical performances and computational advantage of the proposed method over existing alternatives. Our method considerably improves inference efficiency and prediction accuracy at multiple quantile levels compared to several popular competitors under a diverse set of hypothetical truths. Moreover, our model exhibits great adaptation to different correlation strengths and accurately estimates copula parameters. We also introduce an extension based on a $t$ copula process to model heavy tailed, spatially dependent response data (Section \ref{sec:tail}), and demonstrate that the Watanabe-Akaike information criteria \citep{watanabe2010asymptotic} could be used effectively to select between competing spatial joint quantile regression models with different marginal and dependency structures. 

Furthermore, it is demonstrated that in analyzing \pmc~concentration in the northeast US, substantial inferential gains are made by employing a spatial quantile regression analysis over standard spatial random effects methods (Section \ref{sec:case}). The new method offers substantially better fit to the data as measured by hold-out quantile prediction errors, and, true to the promise of QR, estimates predictor effects to vary substantively across \pmc~quantile levels. Particularly, the effects of the percentage of urban land use and nearby \pmc~emission have clear descending and ascending trends as the quantile level increases. These predictors appear to have a different impact on very high levels of \pmc~concentration than on moderate or mild levels.

We end this introduction with the note that the problem of quantile regression under spatial dependency is quite different from the one of spatially varying quantile regression as explored by \cite{reich2011bayesian} and \cite{yang2015quantile}. Statistical analysis in the latter problem fundamentally relies on having independent replications of the response measurement at identical or nearby locations and existing methods essentially employ spatial smoothing of locally estimated quantile regression parameters. More fundamentally, the underlying theory posits that the structural equations representing the response-predictor relation vary across space. In contrast, our approach theorizes that a single quantile regression formulation as in \eqref{equ:lqrmodel} holds globally at all spatial locations, and, the learning of these global model parameters, while adjusting for spatial dependency, is the primary goal of the analysis. However, our model does allow spatial quantile smoothing necessary for {\it infill} prediction at new locations where only the covariates are recorded. Although this type of smoothing has less shape flexibility than what is offered by fully spatially varying approaches, it can still offer a superior performance under moderate spatial variation of predictor effects (Section \ref{sec:discussion}).

\section{Joint spatial quantile regression}
\label{sec:jsqr} 
\subsection{The model}
\label{subsec:model}
Our focus is on analyzing spatial point-referenced data where univariate response data $Y_i$, $i = 1, \ldots, n$, are collected along with associated locations $s_i\in\mathcal{S}\subset\mathbb{R}^r$ and predictors $X_i\in\mathcal{X}\subset\mathbb{R}^p$. %We characterize spatial dependence across observations by allowing underlying quantile levels to be spatially dependent. %This approach is based on a simple yet useful idea that if observations are spatially dependent, the underlying quantile levels are also spatially dependent. 
%This rationale complies with the law of nature. For example, if the temperature at a location is high and reaches above the 80th percentile of the distribution of temperatures across different locations, then a nearby location is also likely to have a high temperature with a high percentile and vice versa. 
The possibility of spatial correlation between nearby observation units implies that $\{U_i=U(s_i)\}_{i=1}^n$ maybe dependent. But they must remain marginally distributed according to $\Un(0,1)$ in view of the generative model \eqref{equ:generative}. By Sklar's theorem \citep{sklar1959fonstions}, the joint distribution of $(U_1,...,U_n)$ must be given by some copula $C(\cdot,\ldots,\cdot)$. Coupled with the marginal model \eqref{equ:generative}, our joint spatial quantile regression model is
\begin{align}
Y_i&=\beta_0(U_i)+X_i^\top\beta(U_i),\quad\quad(U_1,\ldots,U_n)\sim C(u_1,\ldots,u_n),
\label{equ:datagen}
\end{align}
which, again, is a fully generative model for the observed response data that simultaneously addresses possible spatial correlation between these records. 

The beauty of Sklar's theorem is that it allows the marginal distributions and the dependence between the margins to be modeled separately. The framework of \yt~is adopted for inference of marginal quantile functions. Towards a statistical analysis of dependence, it is pragmatic to consider a parametric family of copulas $\{C(u_1,\ldots,u_n |\theta): \theta\in\Theta\}$ indexed by a low dimensional parameter $\theta$ describing spatial dependence across observations.
%and, with $c(\cdot,\ldots,\cdot|\theta)$ denoting the copula density function, the log-likelihood score of model parameters can be expressed as
%\begin{align}
%\ell(\gamma_0,\gamma,\sigma,\omega,\zeta,\theta)=-\sum_{i=1}^n\log\{\dot{\beta}_0(U_i)+X_i^\top\dot{\beta}(U_i)\}+\log c(U_1,\ldots,U_n\mid\theta)
%\label{equ:loglik}
%\end{align}
%where $U_i$ is still found by solving \eqref{equ:taueq}. The log-likelihood can be partitioned into two pieces, one marginal and one copula. The marginal part remains same as the log-likelihood \eqref{equ:indloglik} for independent observations with $(\gamma_0,\gamma,\sigma,\omega,\zeta)$ characterizing the marginal quantile functions. The copula part with parameter $\theta$ describes the spatial dependence across observations.

To appropriately incorporate spatial dependence, a copula family must satisfy the following principles. The dependency induced by the copula should not be affected by the sampling order of observations. The dependency pattern should be invariant regardless of whether a particular sample is included in the analysis or not. These two principles share a similar spirit with the Kolmogorov extension theorem and guarantee the compatibility of performing spatial predictions at new locations. For modeling and computational ease, it is expedient to let the dependency between any two observations depend only on their spatial distance. It is absolutely crucial that as this distance increases, the observations should be less dependent, and near-independence should be realized within the range of the observed spatial domain. %This is a requirement for both copula type and range of copula parameters $\theta$ (Section \ref{subsec:prior}). 
Any hope of reliable inference of the marginal parameters rests on this requirement.

This desiderata rules out the use of many prevalent copula families. For example, multivariate Archimedean copulas have exchangeable dependence structures, that is, all pairwise dependencies are exactly identical. The pairwise correlation induced by the Farlie-Gumbel-Morgenstern copula has a limited range and decreases as sample size increases \citep{mari2001correlation}. The dependency between observations  produced by the structured factor copula model \citep{krupskii2015structured} depends on the order of samples. As the computation complexity of the multivariate $\chi^2$ copula \citep{bardossy2006copula} density is $O(2^n)$, it has limited practical appeal in our context. While Vine copulas are flexible in modeling complex dependence structures, a careful design of the $n(n-1)/2$ bivariate margins would be required to satisfy our desiderata, and the resulting $O(n^2)$ parameters would be extremely challenging to estimate from limited data. 

Although it is challenging to construct a general Vine copula to satisfy our desiderata, an attractive special choice manifests in a Gaussian copula, induced by a stationary Gaussian spatial process. This choice has several additional advantages. The correlation function of the Gaussian spatial process could be used to specify well structured spatial dependency models with only a small number of unknown parameters controlling the smoothness and the decay range of the correlation. The conditional copula distributions and quantile functions, key quantities needed for infill prediction and spatial interpolation, are straightforward to compute (Section \ref{subsec:postinfer}). Equally important, the use of Gaussian process is compatible with the rich literature on spatial modeling and incorporates the popular basic spatial random effects model as a special case \citep[][more details below]{cressie1993statistics, banerjee2008gaussian}.

The Gaussian copula process is specified as follows. Let $U_i=\Phi(Z(s_i))$ where $\Phi(\cdot)$ is the cumulative distribution function (CDF) of $\N(0,1)$. We define
\begin{align}
Z(s_i)=W(s_i)+\varepsilon(s_i),~W(s)\sim\textsf{GP}\left(0,\alpha\rho(s,s';(\nu,\phi))\right),~\varepsilon(s_i)\simiid\N(0,1-\alpha)
\label{equ:gpmodel}
\end{align}
where $\rho(\cdot,\cdot;(\nu,\phi))$ can be any valid correlation function with smoothness parameter $\nu$ and decay parameter $\phi$. We adopt the Mat\'ern correlation function with parametrization
\begin{align*}
\rho_\text{M}(s,s';(\nu,\phi))=\frac{2^{1-\nu}}{\Gamma(\nu)}\left(\sqrt{2\nu}\frac{\|s-s'\|}{\phi}\right)^\nu K_\nu\left(\sqrt{2\nu}\frac{\|s-s'\|}{\phi}\right).
\end{align*}
where $\Gamma(\cdot)$ is the Gamma function and $K_\nu(\cdot)$ is the modified Bessel function of the second kind of order $\nu$.

Our construction is attractive on two accounts. First, the variation in the underlying quantiles is decomposed into two parts in a similar spirit as the basic spatial random effects (BSRE) model \eqref{equ:bsre}. The process $W(s)$ captures structural spatial association while $\varepsilon(s)$ is uncorrelated pure error. The parameter $\alpha\in[0,1]$ determines the proportion of variation that is spatially structured. When $\alpha=1$, no ``nugget'' effect is included; when $\alpha=0$, the model essentially degenerates to independent scenario. As a desirable byproduct, the introduction of $\alpha$ also improves numerical stability. Another compelling feature of our model is that when $\beta_0(u)=\sigma\Phi^{-1}(u)+\tilde{\beta}_0$ and $\beta(u)=\tilde{\beta}$, the model becomes $Y_i=\tilde{\beta}_0+X_i^\top\tilde{\beta}+\tilde{W}(s_i)+\tilde{\varepsilon}(s_i)$ with $\tilde{W}(s)\sim\textsf{GP}\left(0,\alpha\sigma^2\rho(s,s';(\nu,\phi))\right)$ and $\tilde{\varepsilon}(s_i)\simiid\N(0,(1-\alpha)\sigma^2)$. This special formulation coincides with BSRE. 

%Without special notice, we use \JSQR~to denote our joint spatial quantile regression model \eqref{equ:datagen} with Gaussian copula process \eqref{equ:gpmodel}.

Notwithstanding the many methodological and computational advantages, the Gaussian copula is known to be limited in its ability to model extreme tail dependence and may be inadequate for modeling heavy tailed response distributions. To account for such tail dependence and related issues of tail asymmetry, other copula families including a $t$ copula and skew-elliptical copulas could also be considered. Section \ref{sec:tail} offers an in-depth treatment of the issue of heavy tails and extreme tail dependence. 

\subsection{Prior specification}
\label{subsec:prior}
We adopt a semiparametric Bayesian approach for making inference on model parameters $(\beta_0,\beta,\theta)$. Following \yt, the marginal linear QR model \eqref{equ:generative} respecting the non-crossing condition \eqref{equ:monotone} is first reparametrized and then priors are specified under the new parametrization. Let $f_0$ be a probability density function (pdf) on $\mathbb{R}$ and $q_0$ be the corresponding quantile density, that is, the derivative of the quantile function. Let $\tau_0=\int_{-\infty}^0 f_0(z)\d z$. For any non-zero vector $b\in\mathbb{R}^p$, define ``projection radius opposite $b$'' as $a(b,\mathcal{X})=\sup_{x\in\mathcal{X}}\{-x^\top b\}/\|b\|$ and $a(0,\mathcal{X})=1$, where $\|\cdot\|$ is the Euclidean norm. The \yt~parametrization is
\begin{align}
&\beta_0(\tau_0)=\gamma_0,~~\beta(\tau_0)=\gamma\label{equ:base}\\
&\beta_0(\tau)-\beta_0(\tau_0)=\sigma\int_{\zeta(\tau_0)}^{\zeta(\tau)}q_0(u)\d u,~~\tau\in(0,1)\label{equ:intercept}\\
&\beta(\tau)-\beta(\tau_0)=\sigma\int_{\zeta(\tau_0)}^{\zeta(\tau)} \frac{\omega(u)}{a(\omega(u),\mathcal{X})\sqrt{1+\|\omega(u)\|^2}}q_0(u)\d u,~~\tau\in(0,1)\label{equ:slope}
\end{align}
with model parameters $\gamma_0\in\mathbb{R}$, $\gamma\in\mathbb{R}^p$, $\sigma>0$, $\omega:(0,1)\to\mathbb{R}^p$; and $\zeta:[0,1]\to[0,1]$, a differentiable, monotonically increasing bijection, that is, a diffeomorphism. A logistic transformation is used to replace $\zeta(\cdot)$ with an unconstrained function $\omega_0:(0,1)\to \mathbb{R}$
\begin{align*}
\zeta(\tau)=\frac{\int_0^\tau \exp\{\omega_0(u)\}\d u}{\int_0^1 \exp\{\omega_0(u)\}\d u},~~\tau\in(0,1).
\end{align*}
At this point, all function valued parameters $\omega=(\omega_1,...,\omega_p)$ and $\omega_0$ are unconstrained and hence well suited to be assigned Gaussian process priors to induce smoothness regularization. The priors we work with are
\begin{align*}
\omega_j&\sim\textsf{GP}(0,\kappa_j^2\rho_\text{SE}(\cdot,\cdot;\lambda_j))~~j=0,\ldots,p\\
\kappa_j^2&\simiid\textsf{Inv-Ga}(0.1,0.1)~~\lambda_j\sim\pi_\lambda(\lambda_j)\\
(\gamma_0,\gamma,\sigma^2)&\sim \pi(\gamma_0,\gamma,\sigma^2)\propto \frac{1}{\sigma^2}
\end{align*}
where $\rho_\text{SE}(s,s';\lambda)=\exp\left(-\lambda^2\|s-s'\|\right)$ is the square exponential correlation function with rescaling parameter $\lambda$. The prior for $\textsf{Cor}(\omega_j(\tau),\omega_j(\tau+0.1)\mid \lambda_j)=\exp(-0.1^2\lambda_j^2)$ is specified to be $\textsf{Be}(6,4)$. The priors are diffuse and the parameters are to be learned mainly from data. Interested readers may refer to \cite{yang2017joint} for a detailed justification of this choice of priors. 

Because the quantile and quantile density can be efficiently evaluated, standard logistic distribution is selected as a default option of the base distribution $f_0$. When modeling heavy tailed response data, generalized Pareto distribution (for positive response) or $t$ distribution can be used. A comprehensive discussion is delayed until Section \ref{sec:tail}.

To complete our Bayesian estimation formulation, we need to specify prior distributions on the copula parameters $\alpha$, $\nu$ and $\phi$. As a general non-informative prior, $\Un(0,1)$ can be used for $\alpha$. If prior knowledge is available on the value or range of the proportion of spatial correlation, a Beta or a truncated prior may be adopted. The smoothness parameter $\nu$ is typically difficult to estimate from data and often a value of $\nu < 3$ is deemed sufficient for real data analysis \citep{banerjee2014hierarchical}, we fix $\nu$ to be a small, user-defined value depending on specific application. 

More care is needed in choosing a prior for the correlation range parameter $\phi$ in order to adhere to the principles laid out in Section \ref{subsec:model}. Indeed, it is useful to specify a range for $\phi$ from the perspective of \textit{effective range } of spatial correlation, that is, the distance at which the spatial correlation is negligible (typically 0.05). In any particular application, a range for $\phi$ can be determined according to one's belief of lower and upper limits of the effective range. From a methodological perspective, the decay parameter is only weakly identified and an informative prior is needed for satisfactory and reproducible posterior inference and computation \citep{banerjee2008gaussian, banerjee2014hierarchical}. Therefore, we follow the convention in the spatial modeling literature and specify the range of $\phi$ so that the effective range lies between one-fourth and three-fourths of the maximal pairwise distance among all locations \citep{banerjee2014hierarchical}. This choice of range keeps $\phi$ away from 0 and eliminates the identifiability issue that manifests when $\alpha$ approaches 0. 

From a computational perspective, correlation matrix calculation, which requires $O(n^2)$ flops in each MCMC iteration, is expensive. This consideration motivates the use of a discrete uniform prior. Since $\nu$ is fixed, only finitely many possible correlation matrices, induced by the correlation function $\rho$, can be computed and stored before implementing MCMC algorithm.

\section{Posterior computation}
\label{sec:computation}
\subsection{Likelihood evaluation}
\label{subsec:likeval}
By Sklar's theorem, the joint conditional density of responses given predictors can be partitioned into a marginal part and a copula part $\prod_{i=1}^n f_Y(y_i\mid x_i)\times c\left(F_Y(y_1\mid x_1),\ldots,F_Y(y_n\mid x_n)\right)$ where $F_Y$ is the CDF corresponding to pdf $f_Y$ and $c(\cdot,\ldots,\cdot)$ denotes the copula density. A valid specification of $Q_Y(\tau\mid x)$ for all $\tau\in(0,1)$ uniquely determines both $f_Y$ and $F_Y$ \citep{tokdar2012simultaneous} as follows.
\begin{align}
f_Y(y\mid x)=\frac{1}{\frac{\partial}{\partial \tau} Q_Y(\tau\mid x)}\bigg|_{\tau=\tau_x(y)},~~F_Y(y\mid x)=\tau_x(y)
\label{equ:pdf}
\end{align}
where $\tau_x(y)$ solves $Q_Y(\tau\mid x)=y$ in $\tau$. Therefore, based on the model specification \eqref{equ:datagen} and parametrization \eqref{equ:base}--\eqref{equ:slope}, the log-likelihood score of model parameters can be expressed as
\begin{align}
\ell(\gamma_0,\gamma,\sigma,\omega,\zeta,\theta)=-\sum_{i=1}^n\log\{\dot{\beta}_0(U_i)+X_i^\top\dot{\beta}(U_i)\}+\log c(U_1,\ldots,U_n\mid\theta)
\label{equ:loglik}
\end{align}
where $U_i=\tau_{X_i}(Y_i)$ can be obtained by using any univariate root finding algorithm. Such calculation is embarrassingly parallel across observation units. The computation of copula density can be simplified by using spectral decomposition. Due to the discrete nature of the prior for $\phi$, finitely many possible correlation matrices can be calculated before running MCMC.
%Using spectral decomposition $K=\Gamma\Lambda\Gamma^\top$, the correlation matrix $\alpha K+(1-\alpha)I_n$ can be written as $\Gamma(\alpha\Lambda+(1-\alpha)I_n)\Gamma^\top$. Therefore, the logarithm of the Gaussian copula density is 
%\begin{align*}
%\log c(U_1,...,U_n\mid \theta)=-\frac{1}{2}\log |\alpha\Lambda+(1-\alpha)I_n|-\frac{1}{2}Z^\top\Gamma((\alpha\Lambda+(1-\alpha)I_n)^{-1}-I_n)\Gamma^\top Z.
%\end{align*}
%Note that $\alpha\Lambda+(1-\alpha)I_n$ is a diagonal matrix and hence its determinant and inverse are easy to compute. Due to the discreteness nature of the prior for $\phi$, finitely many possible $\Lambda$ and $\Gamma$ can be computed before running MCMC.
\subsection{Inference algorithm}
\label{subsec:algorithm}
Given the likelihood structure \eqref{equ:loglik}, it is natural to employ a Gibbs sampler alternating between updating parameters $(\gamma_0,\gamma,\sigma,\omega,\zeta)$ of marginal part and copula parameters $\theta=(\alpha,\phi)$. However, based on our experience, the scale parameter $\sigma^2$ and $\alpha$ can be highly correlated because both parameters affect the tightness of $U_i$'s. Instead, we find the parametrization $(\sigma_s^2, \sigma_e^2)=(\alpha\sigma^2, (1-\alpha)\sigma^2)$ helps reduce autocorrelation between posterior samples and improves inference efficiency. This parametrization is inspired by noticing that $\sigma_s^2$ and $\sigma_e^2$ are spatial variance and pure error variance respectively when our model degenerates to BSRE \eqref{equ:bsre}. Under this parametrization, we augment the adaptive blocked Metropolis algorithm adopted by \yt~with a learning of copula parameters.

Final details of the algorithm are as follows. First, although it is possible to marginalize $\phi$ out and run MCMC algorithm only on other model parameters, it does not help mixing based on our simulations but increases computation complexity. Therefore we preserve $\phi$ in the cycle of our algorithm. Second, jointly updating $\sigma_s$ and $\sigma_e$ improves MCMC mixing and hence this step is included at end of each iteration. Lastly, the latent process realization $W(s)$ is marginalized out during MCMC runs but can be recovered in post-processing for inference. The algorithm is given below.
\begin{enumerate}
\item Given $\sigma_s$ and $\phi$, update the parameter $(\gamma_0,\gamma,\sigma_e,\omega,\zeta)$.
\item Given $(\gamma_0,\gamma,\sigma_e,\omega,\zeta)$ and $\phi$, update $\sigma_s$.
\item Given the current $U_i$'s extracted from step 2 and $\sigma_s$, sample $\phi$ from its full conditional distribution.
\item Given other model parameters, jointly update $\sigma_s$ and $\sigma_e$.
\end{enumerate}
In the simulation studies presented in Section \ref{sec:experiment}, using the above algorithm with parametrization $(\sigma_s^2,\sigma_e^2)$ increases the effective sample sizes by 27\% and 40\% on average for $p=1$ and $p=7$, compared to using $(\sigma^2,\alpha)$; the mean absolute errors and coverages of regression coefficients are also improved. 

\subsection{Reduced rank approximation}
\label{subsec:approx}
Large matrix factorization and storage can be cumbersome when analyzing spatial datasets with a large number of locations. Although the spectral decomposition is performed before running MCMC, it remains to store all possible correlation matrices and evaluate the copula density multiple times in a single iteration. Reduced rank approximation to Gaussian process can be used to alleviate this computation bottleneck. Prominent examples include nearest-neighbor GP \citep{datta2016hierarchical} and predictive process with pivoting \citep{foster2009stable}.

\section{Spatial smoothing}
\label{sec:spatialsmooth}
\subsection{Prediction of conditional quantile}
\label{subsec:postinfer}
Infill prediction is often one of the major inference objectives in spatial data analysis. Although the model specification \eqref{equ:datagen} appears to target estimating the global quantile predictor effects and spatial dependence separately, our model is able to make infill prediction that adequately accounts for spatial information, thus effectively performing spatial quantile smoothing to new locations where only the covariates have been recorded. Specifically, our model allows for a local adjustment of quantile of response $Y^*$ at $s^*$ via the conditional copula $C(U^*\mid U_1,\ldots,U_n,\theta)$. Denoting its quantile function as $Q_{U^*}(\cdot\mid U_1,\ldots,U_n,\theta)$ and combining the marginal model, the conditional $\tau^*$th quantile of $Y^*$ given $X^*$ is 
\begin{align}
    Q_{Y^*}(\tau^*\mid X^*,s^*,U_1,\ldots,U_n)&=\beta_0(\tau)+X^{*\top}\beta(\tau),~\tau=Q_{U^*}(\tau^*\mid U_1,\ldots,U_n,\theta)\label{equ:conditionalpred}.
\end{align}
It is possible that a new sample is drawn at a location distant from observed locations or is a completely fresh draw considered to be independent with observed samples. In either case, $U^*$ and $(U_1,\ldots,U_n)$ are essentially independent and hence \eqref{equ:conditionalpred} is nearly equal to the marginal quantile $Q_{Y^*}(\tau^*\mid X^*)$. %due to the fact that $\tau\approx\tau^*$. 

When using a Gaussian copula process, the quantile function of the conditional copula can be conveniently calculated. Define $Z=\left(\Phi^{-1}(U_1),\ldots,\Phi^{-1}(U_n)\right)^\top$. Let $K$ be a $n\times n$ matrix with $K_{ij}=\rho(s_i,s_j;(\nu,\phi))$ and $K^*$ be a $n$-dimensional vector with $K^*_i=\rho(s^*,s_i;(\nu,\phi))$. Based on the model specification \eqref{equ:gpmodel}, we have $Z(s^*)\mid (Z,\theta) \sim \N\left(\mu(s^*),\sigma^2(s^*)\right)$ where $\mu(s^*)=\alpha K^{*\top} (\alpha K+(1-\alpha)I_n)^{-1}Z$ and $\sigma^2(s^*)=1-\alpha^2 K^{*\top} (\alpha K+(1-\alpha)I_n)^{-1} K^*$. Therefore, we obtain $Q_{U^*}(\tau^*\mid U_1,\ldots,U_n,\theta) =\Phi(\mu(s^*)+\sigma(s^*)\Phi^{-1}(\tau^*))$.

\subsection{Spatial dependence and spatial variation}
\label{subsec:svc}
It is clear from the above discussion that the quantile function at a new location $s^*$ smoothly varies in $s^*$. This begs the question: could we reinterpret our model as a spatially varying quantile regression model under which $Y_i$'s are conditionally independent? The answer is yes, but in a very limited sense.

%The conditional quantile specified in \eqref{equ:conditionalpred} indeed depends on the location which inspires a spatially varying formulation of our model. %We make a distinction between our approach to quantile regression with spatial dependency adjustment versus spatially varying quantile regression undertaken by, e.g., \cite{reich2011bayesian}. 
Let $V_i=\Phi\left(\varepsilon(s_i)/\sqrt{1-\alpha}\right)$. Given the spatial process realization $W(s)$, the data generating mechanism \eqref{equ:datagen}-\eqref{equ:gpmodel} can be equivalently expressed as
\begin{align}
    Y_i = \beta_0(h_{W,\alpha}(s_i,V_i)) + X_i^\top\beta(h_{W,\alpha}(s_i,V_i)),~V_i\simiid\Un(0,1),1\leqslant i\leqslant n
    \label{equ:svcdatagen}
\end{align}
where $h_{w,a}(s,t)=\Phi(w(s)+\sqrt{1-a}\Phi^{-1}(t))$ with $w:\mathcal{S}\rightarrow\mathbb{R}$, $s\in\mathcal{S}$ and $(a,t)\in(0,1)^2$. Clearly under model \eqref{equ:svcdatagen}, responses $Y_1,\ldots,Y_n$ are conditionally independent given model parameters, spatial process realization $W(s)$ and $X_1,\ldots,X_n$. More importantly, this model admits a quantile function in a spatially varying fashion
\begin{align}
    \tilde{Q}_Y(\tau\mid X,s)=\beta_0\left(h_{W,\alpha}(s,\tau)\right)+X^\top\beta\left(h_{W,\alpha}(s,\tau)\right)
    \label{equ:svcform}
\end{align}
because, the map $t\mapsto h_{w,a}(s,t)$ at each $s\in\mathcal{S}$ is a monotonically increasing diffeomorphism of $(0,1)$ onto itself. Notice that this formulation offers only a limited flexibility in capturing spatial variability since the location $s$ and the quantile level $\tau$ effect the intercept and slope functions through a single, combined input value $h_{w,a}(s,\tau)$.

In contrast, \cite{reich2011bayesian} adopt a fully spatially varying quantile function
\begin{align}
Q_Y(\tau\mid X,s)=\beta_0(s,\tau)+X^\top\beta(s,\tau),
\label{equ:asqrform}
\end{align}
and observations are taken to be conditionally independent. While such a formulation offers much greater shape flexibility in spatial quantile smoothing, it is inadequate for learning a global relationship between the response and the predictors, and, is difficult to estimate. Indeed, the estimation method adopted in \cite{reich2011bayesian} is akin to carrying out a {\it post hoc} Bayesian smoothing of intercept and slope functions estimated locally via the Koenker-Bassett method. Such an approach could be quite useful for analyzing datasets with many repeated observations at each location, even though the two-stage estimation method is likely to offer uncertainty quantifications that are very difficult to interpret. Furthermore, in a simulation study detailed in Supplemental material \ref{subsec:svcillustration}, we observed that \JSQR~offered excellent spatial quantile smoothing, outperforming \cite{reich2011bayesian} when the spatial variation in regression coefficients was moderate. 

%We note here that the formulation \eqref{equ:svcform} is provided purely for the purpose of making connection and distinction to \eqref{equ:asqrform} explicit. The conditional quantile is still given by \eqref{equ:conditionalpred}.

\section{Numerical experiments}
\label{sec:experiment}
We carried out several simulation studies to compare the proposed model against existing methods on inference efficiency and prediction accuracy, and to examine the adaptability of our model to different correlation strengths. In Section \ref{subsec:statperformace}, we demonstrate that in quantile modeling for spatial data, neglecting or inappropriately incorporating spatial dependence results in suboptimal statistical performance by including competing methods that are designed for independent data or offer limited adjustment for spatial dependence. To be comprehensive on simulation design, two examples are included: one with a univariate predictor ($p=1$) and one with multivariate predictors ($p=7$). To investigate robustness of our model against model misspecification, two dependency patterns, one of which deviates from the Gaussian copula process assumption, are considered for each example. In Section \ref{subsec:adaptation}, we illustrate that our model is capable of adapting to different levels of dependence and recovering the true underlying correlation structure by assessing estimation accuracy of copula parameters and induced correlations.

\subsection{Statistical performance assessment}
\label{subsec:statperformace}
\subsubsection{Simulation setup}
\label{subsubsec:simsetup}
The proposed joint spatial quantile regression model (\JSQR\footnote{Without special notice, we use \JSQR~to denote our joint spatial quantile regression model \eqref{equ:datagen} with the Gaussian copula process \eqref{equ:gpmodel}.}) was compared against the classical optimization based method by \cite[\KB]{koenker1978regression}, joint quantile regression model (\JQR) by \yt, Bayesian quantile regression model using asymmetric Laplace error distribution by \cite[\ALDQR]{yu2001bayesian}, and, spatial quantile regression model using asymmetric Laplace process (ALP) by \cite[\ALPQR]{lum2012spatial}. Among these methods, \KB, \JQR~and \ALDQR~do not incorporate spatial dependence. 

Datasets were generated from the marginal models used in \yt~ with spatially dependent underlying quantile levels. The two marginal models were
\begin{description}
    \item[Example 1.] Simple regression with $p=1$:
    \begin{align*}
Q_{Y_i}(\tau\mid X_i)=3(\tau-0.5)\log\frac{1}{\tau(1-\tau)}+4(\tau-0.5)^2\log\frac{1}{\tau(1-\tau)}X_i,~X_i\simiid\Un(-1,1).
\end{align*}
\item[Example 2.] Multiple regression with $p=7$:
\begin{align*}
Q_{Y_i}(\tau\mid X_i)=\beta_0(\tau)+X_i^\top\beta(\tau),~X_i\simiid \Un\left(\{x\in\mathbb{R}^7:\|x\|\leqslant 1\}\right)
\end{align*}
with
\begin{align*}
\beta_0(0.5)&=0,~\beta(0.5)=(0.96,-0.38,0.05,-0.22,-0.80,-0.80,-5.97)^\top\\
\dot{\beta}_0(\tau)&=\frac{1}{\phi(\Phi^{-1}(\tau))},~\dot{\beta}(\tau)=\frac{\dot{\beta}_0(\tau)\nu(\tau)}{\sqrt{1+\|\nu(\tau)\|^2}}\\
\nu_j(\tau)&=\sum_{l=1}^3 a_{lj}\phi(\tau;(l-1)/2,1/9),~~j=1,\ldots,7\\
a&=\left(\begin{array}{ccccccc}
0 & 0 & -3 & -2 & 0 & 5 & -1\\
-3 & 0 & 0 & 2 & 4 & 1 & 0\\
0 & -2 & 2 & 2 & -4 & 0 & 0
\end{array}\right)
\end{align*}
where $\phi(\cdot)$ denotes the pdf of $\textsf{N}(0,1)$. 
\end{description}
Example 1 has an S-shaped intercept and a U-shaped slope function. The configuration in Example 2 defines valid conditional distributions of $Y$ given arbitrary $X$ in the unit ball in $\mathbb{R}^7$ with quantile functions specified above. The base distribution of $Y$ at $X=0$ is $\textsf{N}(0,1)$. %To generate $X$, we sample $Z\sim\textsf{N}(0,I_7)$, $R\sim\textsf{Be}(7,1)$ and set $X=RZ/\|Z\|$. $\beta_0(\tau)$ and $\beta(\tau)$ are numerically evaluated using \textsf{integrate()} function in R.

For each marginal model, we considered two dependence structures. 
\begin{enumerate}[(i).]
\item Asymmetric Laplace copula process \citep{lum2012spatial}:
\begin{gather*}
U_i=F_\text{AL}\left(W(s_i);\tau\right),~W(s_i)=\sqrt{\frac{2\xi_i}{\tau(1-\tau)}}Z(s_i)+\frac{1-2\tau}{\tau(1-\tau)}\xi_i,~\xi_i\simiid\textsf{Exp}(1)\\
Z(s)\sim \textsf{GP}(0,k(s,s')),~k(s,s')=\alpha\rho_\text{M}(s,s';(\nu,\phi))+(1-\alpha)\mathds{1}(s=s')
\end{gather*} 
where $F_{\text{AL}}(\cdot;\tau)$ is the CDF of asymmetric Laplace distribution $\textsf{AL}(\tau)$ with pdf $f_{\text{AL}}(x;\tau)=\tau(1-\tau)\exp\{-\rho_\tau(x)\}$.
%and $\rho_\tau(\varepsilon)=\varepsilon\{\tau-\mathds{1}(\varepsilon < 0)\}$ is the so-called check loss function \citep{koenker1978regression}. 
\item Gaussian copula process: 
\begin{align*}
U_i=\Phi(Z(s_i)),~Z(s)\sim \textsf{GP}(0,k(s,s')),~k(s,s')=\alpha\rho_\text{M}(s,s';(\nu,\phi))+(1-\alpha)\mathds{1}(s=s')
\end{align*}
\end{enumerate}
\cite{kozubowski2000multivariate} show that $W(s_i)$ marginally follows $\textsf{AL}(\tau)$. For each observation, we generated its location $s$ uniformly from $[0,1]^2$. In all simulations, we set $(\alpha,\nu,\phi)=(0.7,2,0.3)$ and $\tau=0.4$. The copula induced by ALP is asymmetric and was used to assess the robustness of \JSQR~under model misspecification. As the asymmetric Laplace copula process was adopted on latent quantile levels instead of random errors, the data generating mechanism here was fundamentally different from that of \ALPQR. Therefore \ALPQR~is misspecified under any quantile level. We generated 100 synthetic datasets under each combination of marginal model and copula model. Each dataset consisted of a training set with $n=200$ and $n=500$ observations for Example 1 and Example 2 respectively and a test set containing 50 observations.

Different methods were compared based on mean absolute errors (MAE) of point estimates of regression coefficients and corresponding coverage probabilities of 95\% confidence (or credible) intervals. For two simulation studies with the Gaussian copula process\footnote{The conditional quantile function $Q_{Y}(\tau\mid X,s,U_1,\ldots,U_n)$ is not analytically available for the asymmetric Laplace copula process.}, we also compared the MAE of predicted conditional quantiles $|Q_{Y}(\tau\mid X,s,U_1,\ldots,U_n)-\hat{Q}_{Y}(\tau\mid X,s,U_1,\ldots,U_n)|$ averaged over all observations in the test set. Here $\hat{Q}_{Y}(\tau\mid X,s,U_1,\ldots,U_n)$ is the predicted conditional quantile function given by different methods. For \KB, \JQR~and \ALDQR, this function is precisely $\hat{Q}_{Y}(\tau\mid X)$. For \ALPQR, we adopted the spatially adjusted quantile function by conditioning on the latent process. All these comparisons were performed at quantile levels $\{0.01,0.05,0.1,\ldots,0.9,0.95,0.99\}$ and averaged over all synthetic datasets under each scenario. For Bayesian methods, posterior means were used as point estimates. The confidence intervals for \KB~were constructed by inverting a rank test proposed by \cite{koenker1994confidence}. 

The algorithm configurations are as follows. The base density function $f_0$ was specified to be the pdf of standard logistic distribution for both \JQR~and \JSQR. We followed the prior specifications for \JSQR~in Section \ref{subsec:prior} with smoothness parameter $\nu=2$ and employed $10$ discrete values for decay parameter $\phi$. \ALDQR~was implemented with default non-informative priors \citep{benoit2017bayesqr}. We implemented \ALPQR~according to the model fitting procedure described in Section 5 and Appendix A in \cite{lum2012spatial}. We found that their method is very sensitive to the prior choice. In all numerical experiments, a $\textsf{Be}(1,6)$ prior was assigned for the proportion of spatial variation $\alpha$ instead of uniform prior stated in their paper to achieve better estimation accuracy. The underlying Gaussian process configuration was kept same as \JSQR. 

\KB, \JQR~and \ALDQR~were implemented using R packages \textsf{quantreg}, \textsf{qrjoint} and \textsf{bayesQR} respectively. \JSQR~and \ALPQR~were implemented in C with R wrapper. We ran all Bayesian methods for 20,000 MCMC iterations with first 10,000 samples as burn-in. We thinned the remaining samples and preserved 500 samples for posterior inference. Except for \ALPQR, we tracked the estimates over a grid of $\tau$ from 0 to 1 with increment 0.01. We use the simulation Example 2 to illustrate the computing time for each algorithm to analyze one dataset. On a desktop with CPU Intel(R) Core(TM) i7-4790 3.60GHz, \JSQR, \JQR~and \ALDQR~took 6.3, 2.6 and 9.2 minutes respectively. \KB~took only 0.07 seconds. \ALPQR~took 36 minutes to perform an analysis at one quantile level and hence took 468 minutes in total for the quantile levels used for summary.
\begin{figure}[h]
\centering
\includegraphics[width=16cm]{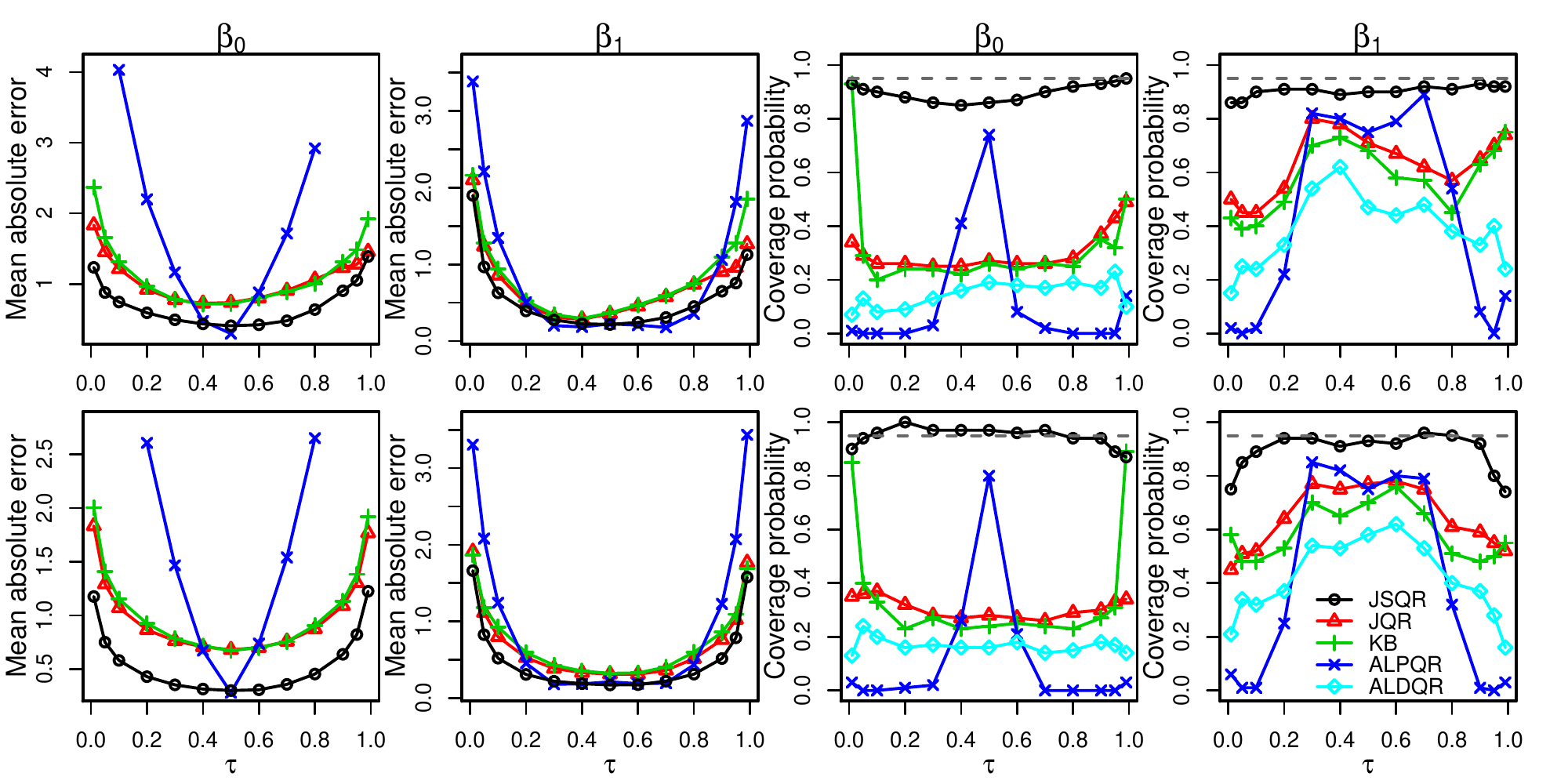} 
\caption{\label{fig:ex1maecp}Inference efficiency of different methods for Example 1 with true generating copula process being the asymmetric Laplace copula process (top row) and the Gaussian copula process (bottom row). In each row, the left two panels present the mean absolute errors of regression coefficients and the right two panels show the coverage probabilities of 95\% confidence (or credible) intervals of regression coefficients at $\tau\in\{0.01,0.05,0.1,\ldots,0.9,0.95,0.99\}$.} %\JSQR: proposed joint spatial quantile regression; \JQR: joint quantile regression by \cite{yang2017joint}; \KB: classical optimization-based method by \cite{koenker1978regression}; \ALPQR: spatial quantile regression using asymmetric Laplace error process by \cite{lum2012spatial}; \ALDQR: quantile regression using asymmetric Laplace error distribution by \cite{yu2001bayesian}.}
\end{figure}

\subsubsection{Analysis of simulation results}
\label{subsubsec:simresult}
\begin{figure}[!t]
\centering
\includegraphics[width=16cm]{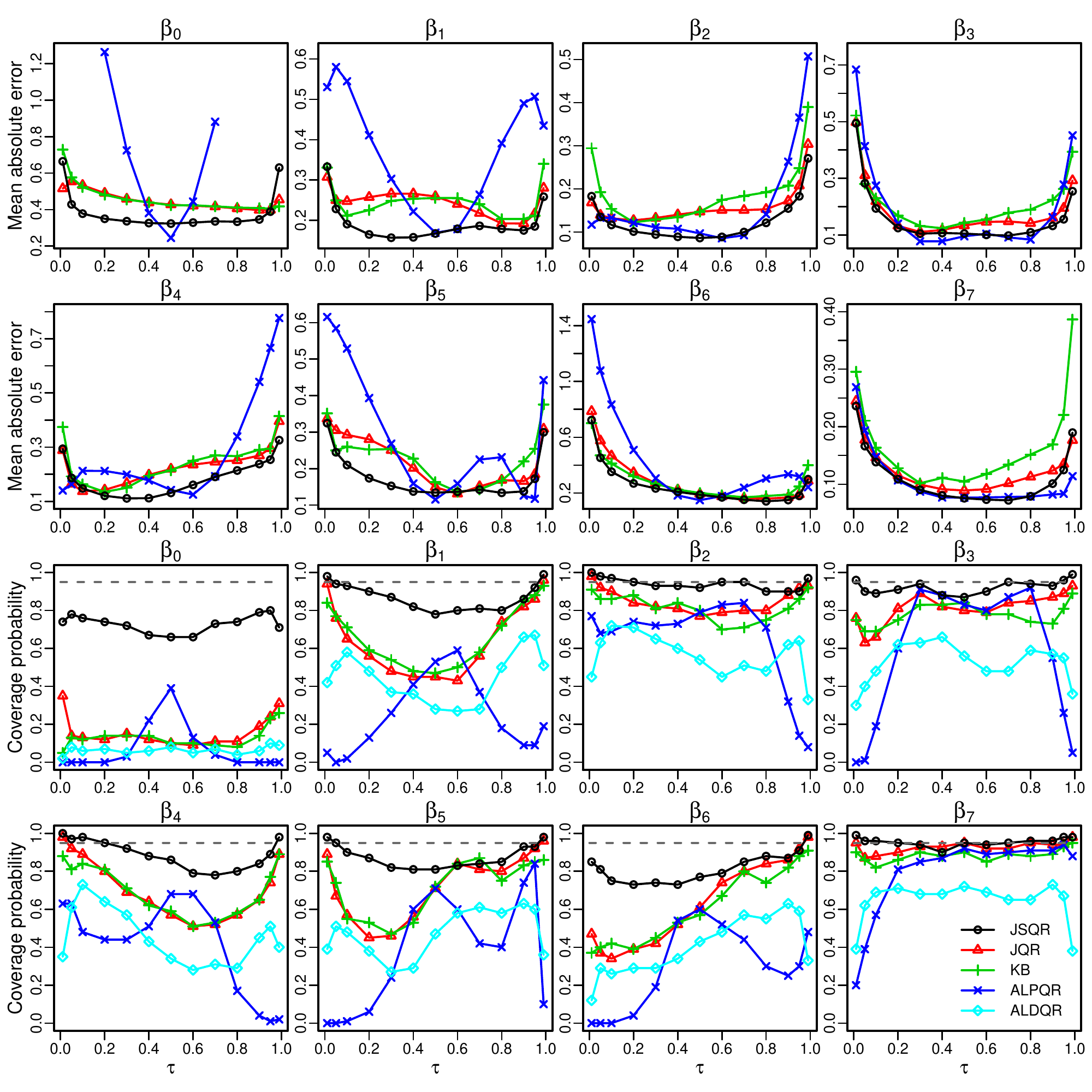}
\caption{\label{fig:ex2alpmaecp}Inference efficiency of different methods for Example 2 with true generating process being asymmetric Laplace process. The top two rows present the mean absolute errors of regression coefficients while the bottom two rows show the coverage probabilities of 95\% confidence (or credible) intervals of regression coefficients at $\tau\in\{0.01,0.05,0.1,\ldots,0.9,0.95,0.99\}$.} %\JSQR: proposed joint spatial quantile regression; \JQR: joint quantile regression by \cite{yang2017joint}; \KB: classical optimization-based method by \cite{koenker1978regression}; \ALPQR: spatial quantile regression using asymmetric Laplace error process by \cite{lum2012spatial}; \ALDQR: quantile regression using asymmetric Laplace error distribution by \cite{yu2001bayesian}.}
\end{figure}

Since the simulation results for Example 2 with two dependence structures were similar, the results for the Gaussian copula process are detailed in Supplemental material \ref{app:ex2gpfig}. According to Figure \ref{fig:ex1maecp}, \ref{fig:ex2alpmaecp} and \ref{fig:ex2gpmaecp}, \JSQR~consistently provided smaller MAEs and better coverages of regression coefficients. The performances of \KB~and \JQR~were similar in general. \JQR~occasionally provided smaller errors ($\beta_7(\tau)$ in Example 2) and better coverages ($\beta_1(\tau)$ in Example 1) than \KB. Since non-informative prior was used, the point estimates of regression coefficients given by \ALDQR~were similar to those by \KB~and hence are not plotted for clarity.

We only present the MAEs of intercept provided by \ALPQR~at quantiles from 0.2 to 0.8 as the estimates were severely biased out of this range. The use of asymmetric Laplace error process distorted the estimation of intercept. Correspondingly, the coverage dropped rapidly at extreme quantiles. In most cases, \ALPQR~poorly estimated regression coefficients at quantile levels out of $(0.4,0.6)$. Moreover, \ALPQR~and \ALDQR~provided lower coverage probabilities than all other methods. The poor statistical performances suggest that simply using the asymmetric Laplace error distribution is inadequate for modeling general distributions. 

\begin{figure}[h]
\centering
\includegraphics[width=12cm]{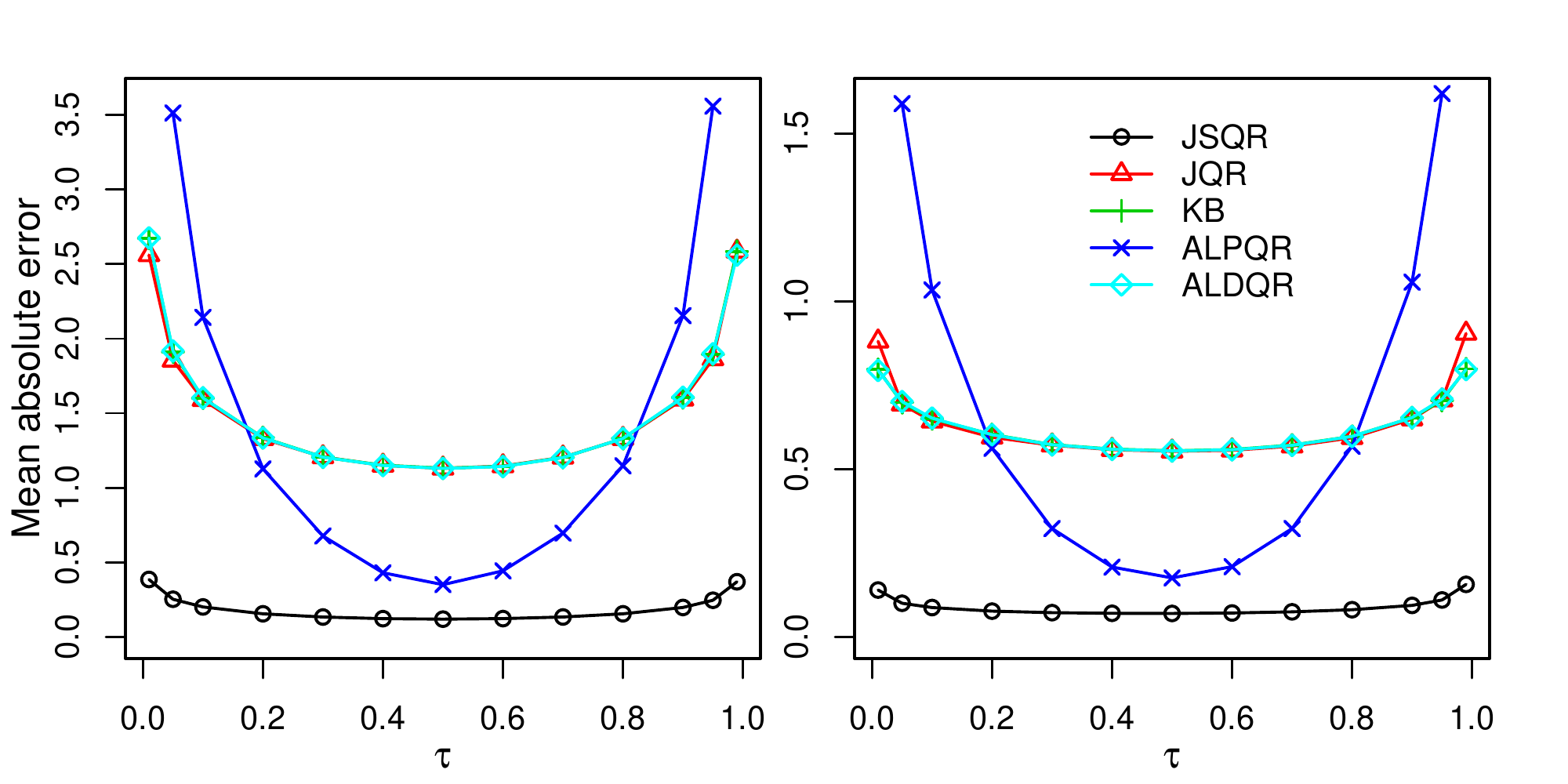} 
\caption{\label{fig:maecondqt}Mean absolute errors of conditional quantile function at $\tau\in\{0.01,0.05,0.1,\ldots,0.9,0.95,0.99\}$ for Example 1 (left) and Example 2 (right) with the Gaussian copula process.} %\JSQR: proposed joint spatial quantile regression; \JQR: joint quantile regression by \cite{yang2017joint}; \KB: classical optimization-based method by \cite{koenker1978regression}; \ALPQR: spatial quantile regression using asymmetric Laplace error process by \cite{lum2012spatial}; \ALDQR: quantile regression using asymmetric Laplace error distribution by \cite{yu2001bayesian}.}
\end{figure}

According to Figure \ref{fig:maecondqt}, the MAEs of conditional quantiles provided by \JSQR~were substantially smaller than those given by other methods. \JQR, \KB~and \ALDQR~provided nearly equal MAEs. The MAEs provided by \ALPQR~were smaller than these three methods at quantiles from 0.2 to 0.8 and were greater than all other methods at extreme quantile levels. The large MAEs given by \ALPQR~at quantiles 0.01 and 0.99 are not presented in the figure.

These empirical evidences positively suggest that the proposed method dominated all other competitors in both inference efficiency and prediction accuracy. Furthermore, our method was robust against copula misspecification. Although \ALPQR~poorly estimated regression coefficients at quantile levels out of $(0.4,0.6)$, this model was able to adjust for spatial dependence (to some extent) when making prediction of conditional quantiles at moderate quantile levels. However, \KB, \JQR~and \ALDQR~failed to account for spatial dependence resulting in poor estimation and prediction at any quantile level.

%Posterior summaries of copula parameters are provided in Table \ref{tab:copulapara}. The estimates of $\phi$ are similar under ALP and GP for both examples and close to the true value 0.3. The estimate of $\alpha$ is more biased when the true generating process is ALP as our model is misspecified. 
%\begin{table}[H]
%    \centering
%    \begin{tabular}{ccccccc}
%    \hline
%    \hline
%    & & \multicolumn{2}{c}{Ex 1, $p=1$} && \multicolumn{2}{c}{Ex 2, $p=7$}\\
%    & & ALP & GP && ALP & GP\\
%    \hline
%    $\alpha$ & & $0.55_{0.10}$ & $0.64_{0.09}$ & & $0.62_{0.08}$ & $0.74_{0.08}$\\
%    $\phi$ & & $0.28_{0.04}$ & $0.28_{0.04}$ & & $0.32_{0.03}$ & $0.32_{0.03}$\\
%    \hline
%    \hline
%    \end{tabular}
%    \caption{\label{tab:copulapara}Posterior summary of copula parameters. For each scenario, we report the average and standard deviation (shown as subscript) of posterior means over 100 synthetic datasets.}
%\end{table}

\subsection{Adaptation to dependence strength}
\label{subsec:adaptation}
To examine the adaptability of our model to different levels of dependence, we assessed the accuracy of estimating the copula parameters and induced pairwise correlations. Specifically, 100 datasets were generated according to the marginal model in Example 1, with spatial dependence generated by the Gaussian copula process but with different and randomly drawn parameter values of $(\alpha, \phi)$. For each dataset, the proportion parameter $\alpha$ was drawn from $\Un(0,1)$ and the decay parameter $\phi$ was randomly drawn from its prior range. Each dataset contained $n=500$ observations. The $\JSQR$ estimation method was employed exactly as in Section \ref{subsec:statperformace}. For each dataset, we estimated $\alpha$, $\phi$ and 10 pairwise correlations $r_{ij}=\alpha\rho(s_i,s_j;(\nu,\phi))$ between 5 randomly selected observed locations, by their respective posterior means. Mean absolute errors of estimates over 100 simulated datasets are reported. We also recorded the coverage probabilities of the respective 95\% credible intervals of these quantities. These results are summarized in Table \ref{tab:copulapara}.

We note that accurately estimating induced pairwise correlations $r_{ij}$ is potentially more important than accurately estimating the copula parameters $\alpha$ and $\phi$. This is because, the induced correlations directly determine the dependence between observations and hence are more critical in adjusting for the spatial dependence toward a more accurate estimation of the marginal QR intercept and slope parameters. Induced correlations are also more identifiable than copula parameters because different combinations of $\alpha$ and $\phi$ may yield similar values of $r_{ij}$. This argument is verified by the results presented in Table \ref{tab:copulapara} which show that our model accurately estimated $r_{ij}$. The 95\% credible interval was narrow and its coverage was close to nominal level. Both absolute errors and length of 95\% credible intervals were smaller than those of $\alpha$ and $\phi$. As the prior range for $\phi$ was around $(0.1,0.4)$, both error of point estimate and uncertainty of $\phi$ were relatively greater than those of $\alpha$ and $r_{ij}$. This is expected as decay parameter is weakly identified. The results suggest that our model is capable of adapting to different levels of dependence and recovering the underlying correlation structure.

\begin{table}[h]
\centering
\begin{tabularx}{0.75\textwidth}{c *{4}{Y}}
\toprule
         &  Absolute error & CP of 95\% CI & Length of 95\% CI\\
         \midrule
        $\alpha$ & $0.05_{0.04}$ & 0.96 & $0.25_{0.11}$\\
        $\phi$ & $0.04_{0.03}$ & 0.95 & $0.17_{0.06}$\\
        $r_{ij}$ & $0.03_{0.04}$ & 0.954 & $0.14_{0.10}$\\
\bottomrule
\end{tabularx}
    \caption{\label{tab:copulapara}Left column: mean absolute errors of posterior means of $\alpha$ and $\phi$ and induced pairwise correlations $r_{ij}=\alpha\rho(s_i,s_j;(\nu,\phi))$. Middle and right columns: coverage probabilities and average lengths of $95\%$ credible intervals of $\alpha$, $\phi$ and $r_{ij}$. Standard deviations are shown as subscripts.}
\end{table}
\section{Modeling heavy tailed response data}
\label{sec:tail}
\subsection{Tail dependence and $t$ copula}
\label{subsec:tcopula}
One of the biggest appeals of QR is that it enables analyzing predictor effects at extreme quantile levels, which could be particularly relevant for analyzing heavy tailed response data. In order to model independent data with heavy tailed response, it is straightforward to adapt \JQR~by adopting a heavy tailed base distribution $f_0$. However, this adaptation of \JSQR~is inadequate for modeling heavy tailed, spatially dependent response because Gaussian copula is \textit{tail independent} \citep{sibuya1960bivariate} and hence would fail to account for the dependence between extreme events. This undesirable copula property may lead to considerably biased predictions of conditional quantile. To this end, it is useful to employ a heavy tailed base distribution coupled with a copula admitting tail dependence in the model. 

Tail dependence quantifies the amount of dependence in the upper or lower tail of a multivariate distribution. Denoting the marginal CDFs of two random variables $X_1$ and $X_2$ as $F_1$ and $F_2$ respectively, the coefficients of upper and lower tail dependence \citep{joe1997multivariate} are defined by the following asymptotic conditional probabilities
\begin{align*}
c_u=\lim_{q\to 1^-}\textsf{P}[X_1>F_1^{-1}(q)\mid X_2>F_2^{-1}(q)],~~c_l=\lim_{q\to 0^+}\textsf{P}[X_1<F_1^{-1}(q)\mid X_2<F_2^{-1}(q)]
\end{align*}
It can be shown that these two coefficients are completely determined by the copula of $(X_1,X_2)$. A copula is said to be upper (lower) tail independent if $c_u=0$ ($c_l=0$), that is, extreme events on univariate margins occur independently. For elliptical copulas including Gaussian copula and $t$ copula, $c_u$ and $c_l$ coincide. 

Since $t$ copula allows for tail dependence \citep{embrechts2001modelling} and can be considered as a generalization of Gaussian copula, we propose using a $t$ copula process to model spatial tail dependence. Following \cite{rasmussen2006gaussian}, we say $f$ is a $t$ process on $\mathcal{X}$ with parameter $\psi>2$, mean function $m(\cdot): \mathcal{X}\rightarrow \mathbb{R}$, covariance function $k(\cdot,\cdot):\mathcal{X}\times\mathcal{X}\rightarrow \mathbb{R}$ if any finite collection of function values $(f(x_1),...,f(x_n))^\top\sim t_n(\psi,m,K)$ where $K_{ij}=k(x_i,x_j)$ and $m_i=m(x_i)$. We denote $f\sim\textsf{TP}(\psi,m,k)$. Here $t_d(\psi,\mu,\Sigma)$ denote a multivariate $t$ distribution on $\mathbb{R}^d$ with location $\mu$, scale matrix $\Sigma$ and degree of freedom $\psi$ with pdf
\begin{align*}
g(z)=\frac{\Gamma\left((\psi+d)/2\right)}{\Gamma(\psi/2)(\psi\pi)^{d/2}|\Sigma|^{1/2}}\left\{1+\frac{(z-\mu)^\top\Sigma^{-1}(z-\mu)}{\psi}\right\}^{-(\psi+d)/2}.
\end{align*}

Let $U_i=T_\psi(Z(s_i))$ where $T_{\psi}$ denote the CDF of $t_1(\psi,0,1)$. We replace \eqref{equ:gpmodel} with
\begin{align}
Z(s)\sim \textsf{TP}(\psi,0,k(s,s')),~k(s,s')=\alpha\rho(s,s';(\nu,\phi))+(1-\alpha)\mathds{1}(s=s').
\label{equ:tpmodel}
\end{align}
A similar additive decomposition can be retrieved via an equivalent Gaussian process mixture representation 
\begin{align*}
Z(s_i)&=W(s_i)+\varepsilon(s_i),~W(s)\sim\textsf{GP}(0,\alpha\rho(s,s';(\nu,\phi))/\varphi)\\
\varepsilon(s_i)&\simiid\textsf{N}(0,(1-\alpha)/\varphi),~\varphi\sim\Gam(\psi/2,\psi/2).
\end{align*}
The amount of tail dependence is controlled by $\psi$ with smaller value inducing greater dependence.  When $\psi$ is moderately large, $t$ process essentially degenerates to Gaussian process. Therefore, we restrict $\psi$ to be in $(2,20)$ and adopt a uniform prior. Prediction of conditional quantile can be performed similarly to that of the Gaussian copula process and is detailed in the Supplemental material \ref{app:tprediction}.

Although $t$ copula admits tail dependence, both $t$ copula and Gaussian copula are not able to accommodate for different lower and upper tail behaviors. To this end, skew-elliptical copulas arisen from skew normal distribution \citep{azzalini1996multivariate} and skew $t$ distribution \citep{azzalini2003distributions} can be considered. %Such copulas capture tail asymmetry through skew parameters and admit conditional Gaussian representations. 
However, we emphasize that tail asymmetry is more relevant in financial applications \citep{mcneil2005quantitative}. Moreover, we already illustrated via simulations in Section \ref{subsec:statperformace} that our model with the Gaussian copula is robust when the true copula for generating data is asymmetric. Therefore, we do not pursue this direction further and refer interested readers to \cite{demarta2005t}  and \cite{smith2012modelling} on the construction of such copulas.

\subsection{Model assessment and comparison}
\label{subsec:modelcomparison}
Because a \JSQR~analysis could be carried out with various combinations of the marginal and dependence models, a natural question arises as to how to compare such competing models and select one with a better model fit. We advocate using the Watanabe-Akaike information criteria \citep[WAIC]{watanabe2010asymptotic, gelman2014understanding} for model comparison and selection. The use of WAIC is particularly attractive in complex Bayesian modeling like ours for at least two reasons. From a theoretical perspective, WAIC has the property of averaging over posterior distributions and is asymptotically equivalent to a  Bayesian leave-one-out cross validation \citep{watanabe2010asymptotic}. From a practical view point, WAIC is computationally attractive as it can be simply calculated by reusing posterior samples. 

For the original \JQR~model for independent data, \cite{cunningham2019vignette} argue that model fit could be assessed directly by evaluating whether the the estimated quantile levels are uniformly distributed, independently of the predictors. They utilize this direct assessment of model fit to underline the usefulness of WAIC in performing model comparison by showing that the two assessments generally agree. Unfortunately, for spatially dependent data, the former diagnostic tool is inappropriate and WAIC is not directly applicable. Nevertheless, as expressed in \eqref{equ:svcdatagen}, observations become independent after conditioning on model parameters $(\gamma_0,\gamma,\sigma,\omega,\zeta,\theta)$ as well as the latent process realization $W(s)$ (and $\varphi$ for the $t$ copula process). Therefore, to apply WAIC, one must also treat $W(s)$ (and $\varphi$) as model parameter(s) and then calculate likelihood values. According to equation \eqref{equ:pdf}, the log-likelihood for $i$th observation with $W(s)$ as an additional parameter is
\begin{align*}
-\log\left\{ \dot{\beta}_0(\Phi(Z(s_i)))+X_i^\top\dot{\beta}(\Phi(Z(s_i)))\right\}-\log \dot{h}_{W,\alpha}(s_i,V_i)
\end{align*}
where 
\begin{align*}
\dot{h}_{W,\alpha}(s_i,V_i)=\frac{\sqrt{1-\alpha}\phi(Z(s_i))}{\phi(\Phi^{-1}(V_i))}.
\end{align*}
The process realization can be recovered according to its posterior
\begin{align*}
    W(s)\mid Z(s),\theta\sim\textsf{N}\left(\frac{1}{1-\alpha}\left(\frac{1}{\alpha}K^{-1}+\frac{1}{1-\alpha}I_n\right)^{-1}Z(s), \left(\frac{1}{\alpha}K^{-1}+\frac{1}{1-\alpha}I_n\right)^{-1}\right)
\end{align*}
and then $V_i=\Phi\left(\left(Z(s_i)-W(s_i)\right)/\sqrt{1-\alpha}\right)$. The derivation for $t$ copula process is similar and is detailed in Supplemental material \ref{app:loglik}.

We highlight model comparisons between the original \JQR~model and the proposed model with the Gaussian copula process and the $t$ copula process. By contrasting these three nested models, we are able to determine whether incorporating spatial dependence and whether using the $t$ copula process result in model improvement. We present the results of model comparison using the second version of WAIC in \cite{gelman2014understanding} for both simulated and real datasets in Section \ref{subsec:modelcompwaic} and Section \ref{sec:case}, respectively. Though the focus here is on the comparison of the proposed models with different combinations of base distribution and copula, WAIC can also be employed to select predictors.

\subsection{Illustration: model comparison using WAIC}
\label{subsec:modelcompwaic}
A simulation study is presented to illustrate the improvement of prediction accuracy offered by the $t$ copula process when analyzing heavy tailed response data and the effectiveness of WAIC on discriminating joint QR models with different dependency structures. We compared \JQR, \JSQR~with the Gaussian copula process (\JSQR-GP) and \JSQR~with the $t$ copula process (\JSQR-TP). All three models adopted $t$ base distribution with degree of freedom to be learned from data. We simulated synthetic datasets with the marginal model
\begin{align*}
Q_{Y_i}(\tau\mid X_i)&=\beta_0(\tau)+\beta(\tau)X_i,~X_i\simiid \Un\left(-1,1\right)
\end{align*}
where 
\begin{align*}
\beta_0(0.5)&=0,~\beta(0.5)=0\\
\dot{\beta}_0(\tau)&=\frac{3}{t_3(T_3^{-1}(\tau),0,1)},~\dot{\beta}(\tau)=\frac{\dot{\beta}_0(\tau)\nu(\tau)}{\sqrt{1+\|\nu(\tau)\|^2}},~\nu(\tau)=3\left(\tau-0.5\right)
\end{align*}
Under this marginal model, three dependency structures were considered: (i) independence, (ii) the Gaussian copula process with $(\alpha,\nu,\phi)=(0.7,2,0.3)$, and, (iii) the $t$ copula process with $(\alpha,\nu,\phi,\psi)=(0.7,2,0.3,3)$. Therefore, each generating model corresponded to one of the models being compared; other two models were either inadequate or redundant. In each scenario, 100 synthetic datasets were generated where each dataset consisted a training set and a test set with 500 and 50 observations respectively.

\begin{figure}[h]
\centering
\includegraphics[width=16cm]{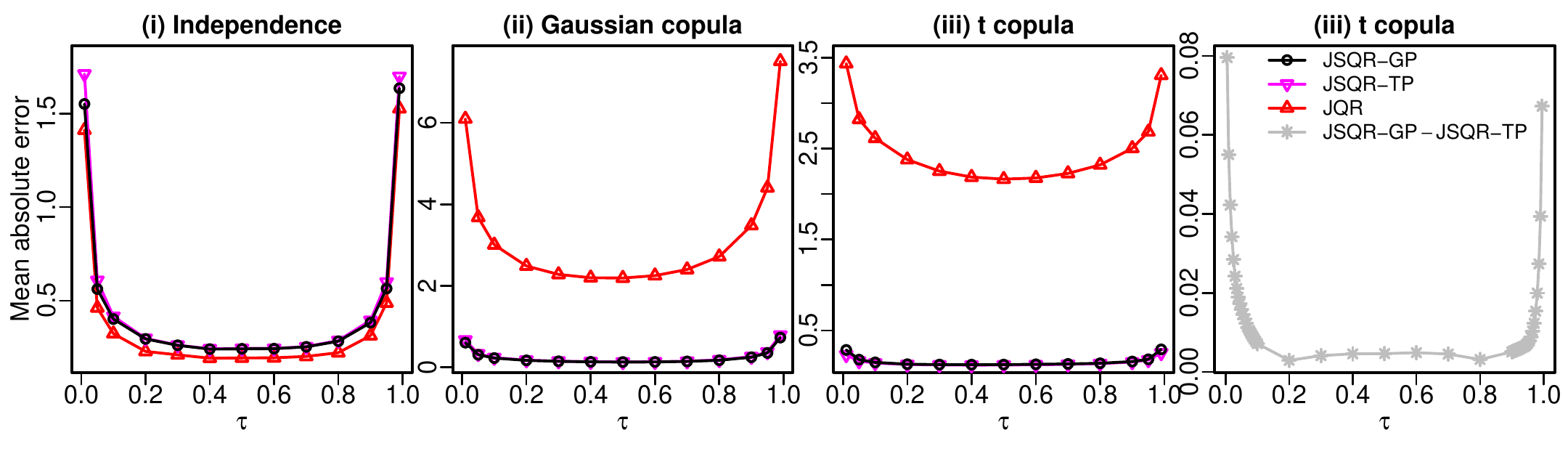}
\caption{\label{fig:waic}First three panels: mean absolute errors of predicted conditional quantiles at $\tau\in\{0.01,0.05,0.1,\ldots,0.9,0.95,0.99\}$. The fourth panel: differences in mean absolute errors of predicted conditional quantiles provided by \JSQR-GP and \JSQR-TP at $\tau\in\{0.005,0.01,\ldots,0.095,0.1,\ldots,0.9,0.905,\ldots,0.990,0.995\}$ for scenario (iii).}
\end{figure}

The true model was chosen as base model in each scenario. The differences in WAIC of other two models and that of the base model are summarized in Table \ref{tab:waicdiff}. The mean absolute errors of predicted condition quantiles were computed (Figure \ref{fig:waic}) over observations in the test set at quantile levels $\{0.01,0.05,0.1,\ldots,0.9,0.95,0.99\}$. We highlight the differences of MAE of conditional quantiles between \JSQR-GP and \JSQR-TP at extreme quantile levels $\{0.005,0.01,\ldots,0.095,0.1\}$ and $\{0.9,0.905,\ldots,0.990,0.995\}$ in scenario (iii). The times one model had the smallest WAIC in each scenario are reported in Table \ref{tab:waicbest}.
\begin{table}[h]
\centering
\begin{tabularx}{\textwidth}{c *{6}{Y}}
\toprule
 & \multicolumn{2}{c}{(i) JQR: $2813_{48}$}   & \multicolumn{2}{c}{(ii) JSQR-GP: $2265_{113}$} & \multicolumn{2}{c}{(iii) JSQR-TP: $2080_{470}$}\\
\cmidrule(lr){2-3} \cmidrule(l){4-5} \cmidrule(l){6-7}
  & JSQR-GP & JSQR-TP & \JQR & JSQR-TP  & \JQR & JSQR-GP \\
\midrule
median & 2.3 & 2.6 & 443.3 & 0.5 & 392.9 & 1.3\\
25\% quantile & 1.0 & 0.9 & 340.0 & -1.0 & 311.5 & 0.1\\
75\% quantile & 3.2 & 4.0 & 522.2 & 1.7 & 498.5 & 2.7\\
\bottomrule
\end{tabularx}
\caption{\label{tab:waicdiff}Quartiles of differences in WAIC of other two models and that of the true model. The mean and standard deviation of WAIC of the true model are reported in the titles.}
\end{table}

%\begin{center}
%\begin{minipage}{0.75\linewidth}
%\begin{table}[H]
%\centering
%\begin{tabularx}{\textwidth}{c *{9}{Y}}
%\toprule
% & \multicolumn{3}{c}{(i) JQR: $2813_{48}$}   & \multicolumn{3}{c}{(ii) JSQR-GP: $2265_{113}$} & \multicolumn{3}{c}{(iii) JSQR-TP: $2080_{470}$}\\
%\cmidrule(lr){2-4} \cmidrule(l){5-7} \cmidrule(l){8-10}
%  & $Q_2$ & $Q_1$ & $Q_3$ & $Q_2$ & $Q_1$ & $Q_3$  & $Q_2$ & $Q_1$ & $Q_3$ \\
%\midrule
%\JQR &   & & & 443.3 & 340.0 & 522.2 & 392.9 & 311.5 & 498.5\\
%      \JSQR-GP & 2.3 & 1.0 & 3.2 &  &  & & 0.1 & 1.3 & 2.7\\
%      \JSQR-TP & 2.6 & 0.9 & 4.0 & -1.0 & 0.5 & 1.7 & & &\\
%\bottomrule
%\end{tabularx}
%\caption{sddfd}
%\end{table}
%\end{minipage}
%\end{center}

According to Table \ref{tab:waicdiff}, \JSQR-GP and \JSQR-TP generally had comparable WAICs because they provided similar model fits if the data only contains a few extreme observations. Due to the penalization of more effective model parameters, the WAICs provided by two \JSQR~models were just slightly larger than \JQR~when observations were independent. The difference was much larger when spatial dependence presented because of a lack of fit in the \JQR~model. According to the last panel in Figure \ref{fig:waic}, the Gaussian copula was indeed inadequate to capture the dependence at extreme quantile levels compared to the $t$ copula when data was generated from a heavy tailed marginal distribution with a tail dependent copula.

We observed a clear consistency between WAIC and MAE of predicted conditional quantile. The differences in WAIC and MAE between any two models were small or large simultaneously. Also, the base models provided the smallest WAICs for most of datasets in all three scenarios according to Table \ref{tab:waicbest}. These evidence suggest that WAIC accurately reflects the predictive performance of a model and is suitable for comparing joint QR models with different dependency structures.
\begin{table}[h]
\centering
\begin{tabularx}{0.4\textwidth}{c *{6}{Y}}
\toprule
 & & i & ii & iii\\
    \midrule
    \JQR & & 77 & 0 & 0\\
    \JSQR-GP & & 12 & 56 & 24\\
    \JSQR-TP & & 11 & 44 & 76\\
\bottomrule
\end{tabularx}
    \caption{\label{tab:waicbest}The percentage one model provides the smallest WAIC.}
\end{table}

\section{Analysis of \pmc~concentration data}
\label{sec:case}
\begin{figure}[h]
\centering
\includegraphics[width=16cm]{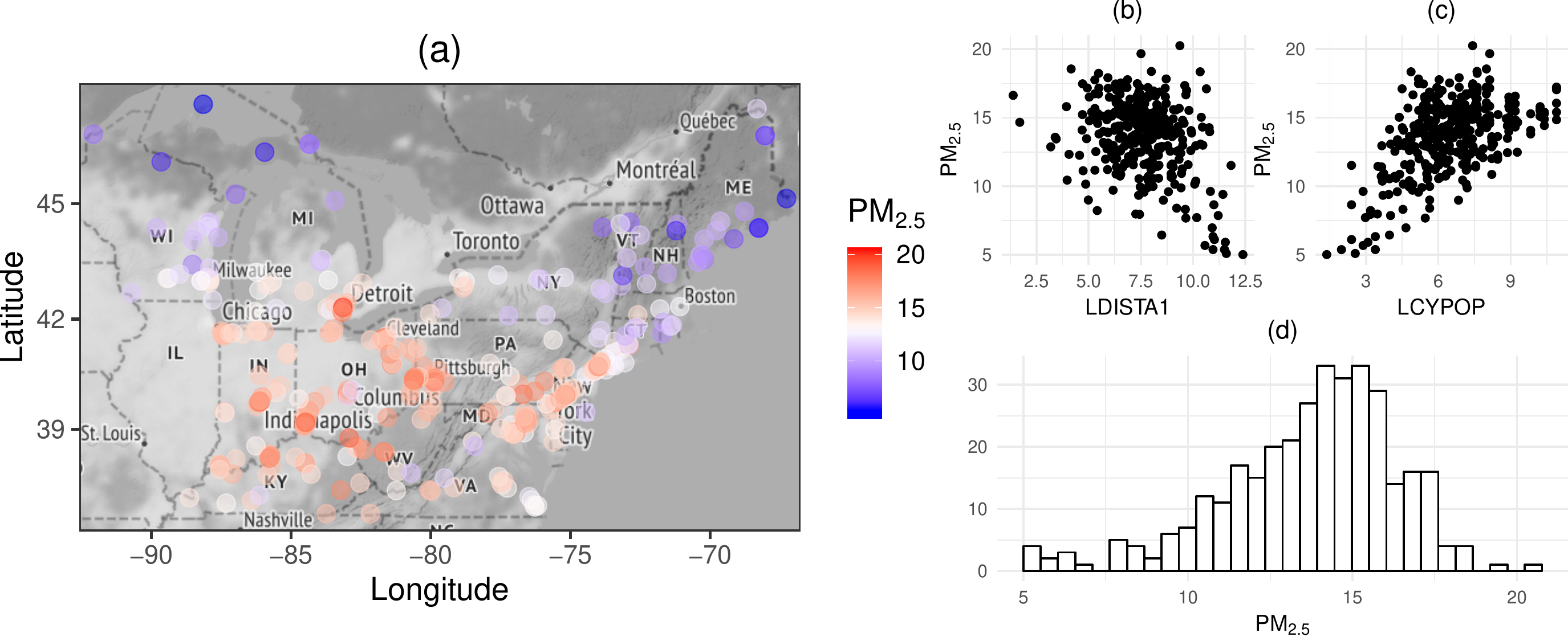}
\caption{\label{fig:datavis}Data visualization. (a) \pmc~concentrations ($\mu \text{g/m}^3$) at monitoring stations across northeastern United States; (b) \pmc~concentrations against \textsf{LDISTA1}; (c) \pmc~concentrations against \textsf{LCYPOP}; (d) Histogram of \pmc~concentrations.}
\end{figure}

Numerous epidemiological studies have proved that both short-term and chronic exposures to fine particulate matter with diameter less than 2.5 $\mu$m are positively associated with the risk for lung cancer and cardiovascular disease morbidity and mortality \citep{dockery1993association, pope2006health, brook2010particulate}. Various regression models, e.g., generalized additive models \citep{yanosky2008predicting}, Cox hazard models \citep{pope1995particulate,jerrett2005spatial} and conditional autoregressive models \citep{paciorek2013spatial}, have been proposed to predict \pmc~concentration. These methods focus only on the response mean prediction. But, arguably, predicting high response quantiles is of a greater concern here because high \pmc~concentration is more harmful as acute exposure could trigger cardiovascular morbidity \citep{bell2005meta}. Moreover, as manifest in Figure \ref{fig:datavis}, the relationship between \pmc~and some important predictors (detailed below) appears more complex than a homoskedastic linear dependence, and, the distribution of \pmc~is skewed. Therefore, classical mean regression analyses are inadequate. Quantile regression has been recently employed to analyze \pmc~\citep{barmpadimos2012one, porter2015investigating}. However, analyses based on \KB~estimates clearly overlook the spatial dependence of data from different monitoring stations as presented in Figure \ref{fig:datavis}(a).

%A case study on particulate matter 2.5 ($\text{PM}_{2.5}$) concentration is presented to demonstrate the potential of the proposed model in real application. In Section \ref{subsec:predper}, we illustrate the superior prediction performance of the proposed model with base distribution and copula selected using WAIC. Insightful quantile predictor effects offered by our model are identified and summarized in Section \ref{subsec:quanpredeff}. A comparison with the constant predictor effects provided by the basic spatial random effects model is included.  

%We start with a brief introduction to the background and then motivate the use of \JSQR~for modeling \pmc~data. 

We reanalyzed data\footnote{The dataset is available at \url{https://www.stat.berkeley.edu/users/paciorek/code/ejs/}.} on \pmc~reported in \cite{paciorek2013spatial} using \JSQR~with a logistic base and the $t$ copula process (\JSQR-TP1), and, \JSQR~with a $t$ base and the $t$ copula process (\JSQR-TP2) in addition to the methods in Section \ref{subsec:statperformace}. The dataset contained averaged daily \pmc~concentrations (shown in Figure \ref{fig:datavis}) from 2001 to 2002 monitored at 339 stations in the northeastern United States. Many geographical and meteorological covariates were recorded. According to \cite{yanosky2008spatio} and \cite{paciorek2009practical}, the following 5 covariates have been shown to have significant associations with \pmc~or improve predictions and hence were included into our analysis: (i) logarithm of population density at county level (\textsf{LCYPOP}), (ii) logarithm of distance to A1 class roads (primary roads, typically interstates) (\textsf{LDISTA1}), (iii) proportion of urban land use within 1 kilometer of the station location (\textsf{URB}), (iv) logarithm of \pmc~emissions within a 10 kilometer buffer (\textsf{L25E10}), and, (v) elevation of the station (\textsf{ELEV}). We refer interested readers to \cite{yanosky2008spatio} for a detailed description of the dataset.

\subsection{Assessment of infill quantile prediction}
\label{subsec:predper}

We performed a 10-fold validation study to assess how well different spatial and non-spatial quantile regression methods performed in predicting response quantiles at held out locations. In each fold, the data was randomly partitioned into a training set and a test set with 270 (about 80\%) and 69 observations respectively. The Euclidean distances between the monitoring stations were calculated. All prior specifications and MCMC settings were kept same as those in Section \ref{sec:experiment}. The accuracy of quantile prediction at a level $\tau$ was evaluated by the check loss $\rho_\tau(Y_i-\hat{Q}_{Y_i}(\tau\mid X_i,s_i,U_1,\ldots,U_n))$ averaged over all observations in both training and test sets. Evaluations were done for $\tau \in \{0.01,0.05,0.1,\ldots,0.9,0.95,0.99\}$ serving as a representative set of quantile levels. 

\begin{figure}[h]
\centering
\includegraphics[width=12cm]{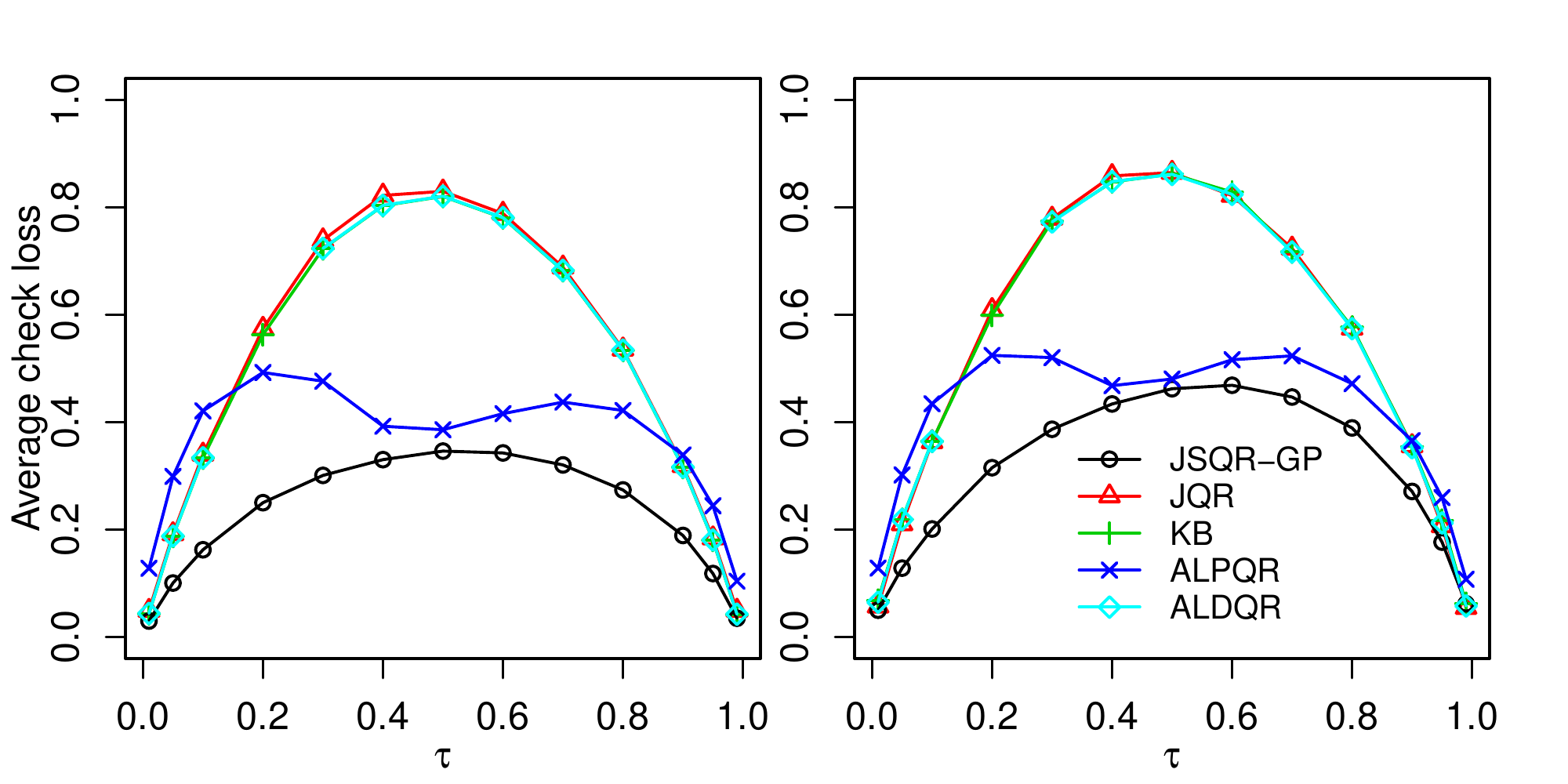}
\caption{\label{fig:pmcheckloss}Average check loss (left: training set; right: test set) at $\tau\in\{0.01,0.05,0.1,\ldots,0.9,0.95,0.99\}$ on \pmc~concentration dataset. The curves for the three \JSQR~models were visually indistinguishable and hence we only present the results provided by \JSQR-GP. 
%\JSQR-GP: proposed joint spatial quantile regression with Gaussian copula process; \JQR: joint quantile regression by \cite{yang2017joint}; \KB: classical optimization-based method by \cite{koenker1978regression}; \ALPQR: spatial quantile regression using asymmetric Laplace error process by \cite{lum2012spatial}; \ALDQR: quantile regression using asymmetric Laplace error distribution by \cite{yu2001bayesian}.}
}
\end{figure}

Results are summarized in Figure \ref{fig:pmcheckloss}. For all methods,  the average check loss on the test set was slightly larger than that on the training set. The three \JSQR~models were comparable to each other and offered the smallest averaged loss consistently at all quantile levels on both training and test sets. The losses given by \JQR, \KB~and \ALDQR~were similar and typically the largest except for extreme quantile levels. \ALPQR~had smaller losses than the three non-spatial methods from $\tau=0.2$ to 0.8 but provided larger losses out of this range. Overall, these results are consistent with our findings in Section \ref{subsec:statperformace} and strongly indicate the necessity of adjusting for spatial dependence toward a more accurate prediction of \pmc~concentration.

The WAICs provided by the four joint QR models (Table \ref{tab:pmwaic}) lends further support for incorporating spatial dependence into a quantile regression of \pmc~concentration. Since the three \JSQR~models offered nearly identical averaged held-out losses, and \JSQR-GP is a simpler model with a slightly smaller average WAIC, we adopted \JSQR-GP for all subsequent analyses.

\begin{table}[h]
\centering
\begin{tabularx}{0.7\textwidth}{c *{4}{Y}}
\toprule
         &  \JQR & JSQR-GP & JSQR-TP1 & JSQR-TP2\\
         \midrule
         WAIC &  $1124_{14}$ & $808_{15}$ & $810_{14}$ & $810_{16}$\\
\bottomrule
\end{tabularx}
    \caption{\label{tab:pmwaic}Average WAIC across 10 training sets with standard deviation shown as subscripts. %\JSQR-GP: proposed joint spatial quantile regression with logistic base and Gaussian copula process; \JQR: joint quantile regression by \cite{yang2017joint} with logistic base; \JSQR-TP1: proposed joint spatial quantile regression with logistic base and $t$ copula process; \JSQR-TP2: proposed joint spatial quantile regression with $t$ base and $t$ copula process.}
    }
\end{table}
 
\subsection{Heterogeneous predictor effect on response quantiles}
\label{subsec:quanpredeff}
To understand how predictors may impact \pmc~concentration heterogeneously across different quantile levels, we refitted the entire dataset with \JSQR-GP and compared parameter estimates against those from the basic spatial random effects model which only allows making inference on mean predictor effects. To recall, the basic spatial random effects model \citep{cressie1993statistics} is
\begin{equation}
Y_i=\beta_0+X_i^\top\beta+W(s_i)+\varepsilon(s_i).
\label{equ:bsre}
\end{equation}
where $W(s)\sim\textsf{GP}\left(0,\sigma_s^2\rho(s,s';\theta)\right)$ can be considered as spatial random effects and $\varepsilon(s)\sim\N(0,\sigma_e^2)$ is independent pure error. %Since $\textsf{E}[Y_i\mid X_i]=\beta_0+X_i^\top\beta$, 
To contrast with \JSQR, notice that the conditional response quantiles under \BSRE~are given by $Q_{Y_i}(\tau\mid X_i)=\beta_0+\sigma_e\Phi^{-1}(\tau)+X_i^\top\beta$.
%Therefore, the intercept $\beta_0+\sigma_e\Phi^{-1}(\tau)$ is varying across different quantiles while slope $\beta$ is invariant.

\begin{table}[h]
\centering
\begin{tabularx}{0.5\textwidth}{c *{2}{Y}}
\toprule
         &  $\alpha~(\sigma_s^2/(\sigma_s^2+\sigma_e^2))$ & $\phi$\\
        \midrule
         \JSQR &   $0.86_{0.04}$ & $0.49_{0.07}$\\
         \BSRE &   $0.81_{0.06}$ & $0.43_{0.08}$\\
\bottomrule
\end{tabularx}
    \caption{\label{tab:pmcopulapara}Posterior means of proportion of spatial variation and decay parameter with standard deviations shown as subscripts.}
\end{table}

\begin{figure}[h]
\centering
\includegraphics[width=16cm]{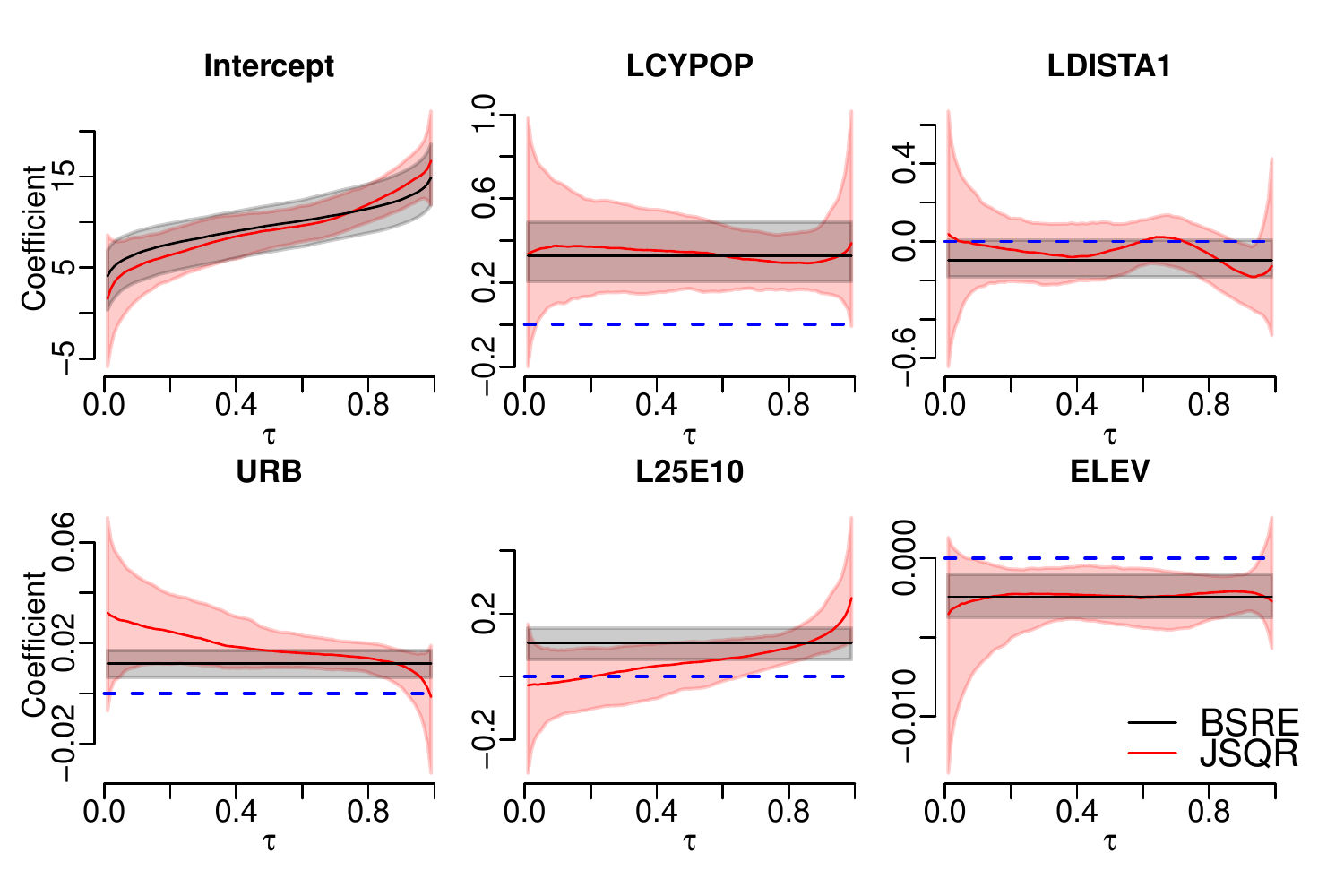}
\caption{\label{fig:pmcoef}Regression coefficients estimated by \BSRE~and \JSQR~along with 95\% credible bands on \pmc~concentration dataset.}
\end{figure}

%The whole dataset was used to fit both models. 
We fitted \BSRE~using the R package \textsf{spBayes}. The posterior means of the proportion of spatial variation and the decay parameter of the Gaussian process (Table \ref{tab:pmcopulapara}) provided by both models suggest strong spatial correlation. The posterior median of regression coefficients along with 95\% credible bands over $\tau\in\{0.01,0.02,\ldots,0.99\}$ are presented in Figure \ref{fig:pmcoef}. \BSRE~and \JSQR~provided similar median and interval estimates for the intercept function across all quantiles. The two models exhibited agreement to some extent on the effects of population density (\textsf{LCYPOP}) and elevation (\textsf{ELEV}) as they were relatively stable across different quantile levels. The sign of their regression coefficients were intuitive and consistent with existing literature \citep{yanosky2008predicting}. Given other predictors in the model, both models suggested that the effect of distance between monitoring station and A1 class roads (\textsf{LDISTA1}) was not significant as the 95\% credible interval contains 0. 

%The intervals reported by \JSQR~at quantile levels from 0 to 0.8 also contains 0. However, at higher quantile levels, the estimate of regression coefficient given by \JSQR~is negative and decreasing indicating that given other predictors, the distance between the monitoring station is negatively associated with \pmc~concentration at high quantile levels.

\begin{table}[h]
\centering
\begin{tabularx}{0.4\textwidth}{c *{1}{Y}}
\toprule
 & $\beta(0.9)-\beta(0.1)$ \\
\midrule
\textsf{URB} & $-0.02~(-0.04,0.00)$\\
\textsf{L25E10} & $0.15~(0.03,0.29)$\\
\bottomrule
\end{tabularx}
\caption{\label{tab:diffeffect}Posterior median and 95\% credible interval of predictor differential effect provided by \JSQR~at $\tau=0.1$ and $0.9$.}
\end{table}
The benefit of \JSQR~over \BSRE~is apparent in the estimates of regression coefficients of the percentage of urban land use (\textsf{URB}) and \pmc~emission within 10 kilometers (\textsf{L25E10}). While \BSRE~estimated the effects of both these predictors to be significant and positive, it was unable to capture further nuances of how their effects vary across quantile levels. In contrast, \JSQR~estimated \textsf{URB} to have a more severe impact on the low and the middle quantile levels, than on the extremely high levels. It also estimated \textsf{L25E10} to have no effects in the low and middle quantile levels, but a fairly pronounced positive effect on the high and extremely high levels. Such differential effects were significant (Table \ref{tab:diffeffect}). These estimates could be interpreted as saying that while higher urban land use results in an increased baseline \pmc~concentration, it is the amount of local \pmc~emission that largely determines the highest \pmc~concentration levels. A rise of \pmc~emission by 2 times within a 10km radius corresponds to an increase of 0.09$\mu \text{g/m}^3$ of the 90th percentile of \pmc~concentration. %[\textcolor{red}{\bf Need actual numbers keeping in mind different units for the two variables. I believe the response is a ``rate'' or density whereas local emission is measured in total volume. Also, L25E10 is measured in log scale}].

\section{Concluding remarks}
\label{sec:discussion}
In this work, we have generalized the joint quantile regression model of \yt~ for analyzing point-referenced spatial data by incorporating spatial dependence via a copula process on underlying quantile levels of observations. The proposed model is versatile; the Gaussian copula process is used for a general purpose while the $t$ copula process can be adopted for modeling heavy tailed response data. A Bayesian semiparametric approach is introduced to perform inference of model parameters. Using posterior samples, model comparison can be effectively performed using WAIC. Extensive empirical evidences are provided to show that the proposed approach dominates existing methods with enhanced statistical performances and reduced computational complexity. 

From a theoretical perspective, the existence of dependence between observations makes it quite challenging to study posterior consistency properties of the proposed Bayesian estimation method. However, we note that the conditional independence implied by the formulation \eqref{equ:svcdatagen} offers a possible path to addressing this challenge under an infill asymptotic setting where $W$ is an additional infinite dimensional parameter to be estimated along with $\beta_0(\cdot)$ and $\beta(\cdot)$. This will be pursued in a future work.

We also note that appropriately chosen copulas could be used to address various other kinds of dependency within a joint quantile regression model, with applications to time series data, longitudinal data, or data where the response is vector valued. 
Such developments will crucially depend on the choice of the copula families that adequately address relevant aspects of statistical estimation and computing. Such efforts are currently underway.

\section*{Acknowledgement}
This research is supported by NSF grant DMS-1613173.

\bibliographystyle{ba}
\bibliography{sample}

%\appendix
%\appendix
%\section*{Supplementary Material}
%\addcontentsline{toc}{section}{Appendices}
%\renewcommand{\thesubsection}{\Alph{subsection}}
\numberwithin{equation}{subsection}

\appendix
\addcontentsline{tic}{section}{Supplemental material}
\section*{Supplemental material}
\renewcommand{\thesubsection}{\Alph{subsection}}

\subsection{Additional simulation results for Example 2 in Section \ref{subsec:statperformace}}
\label{app:ex2gpfig}
\begin{figure}[H]
\centering
\includegraphics[width=16cm]{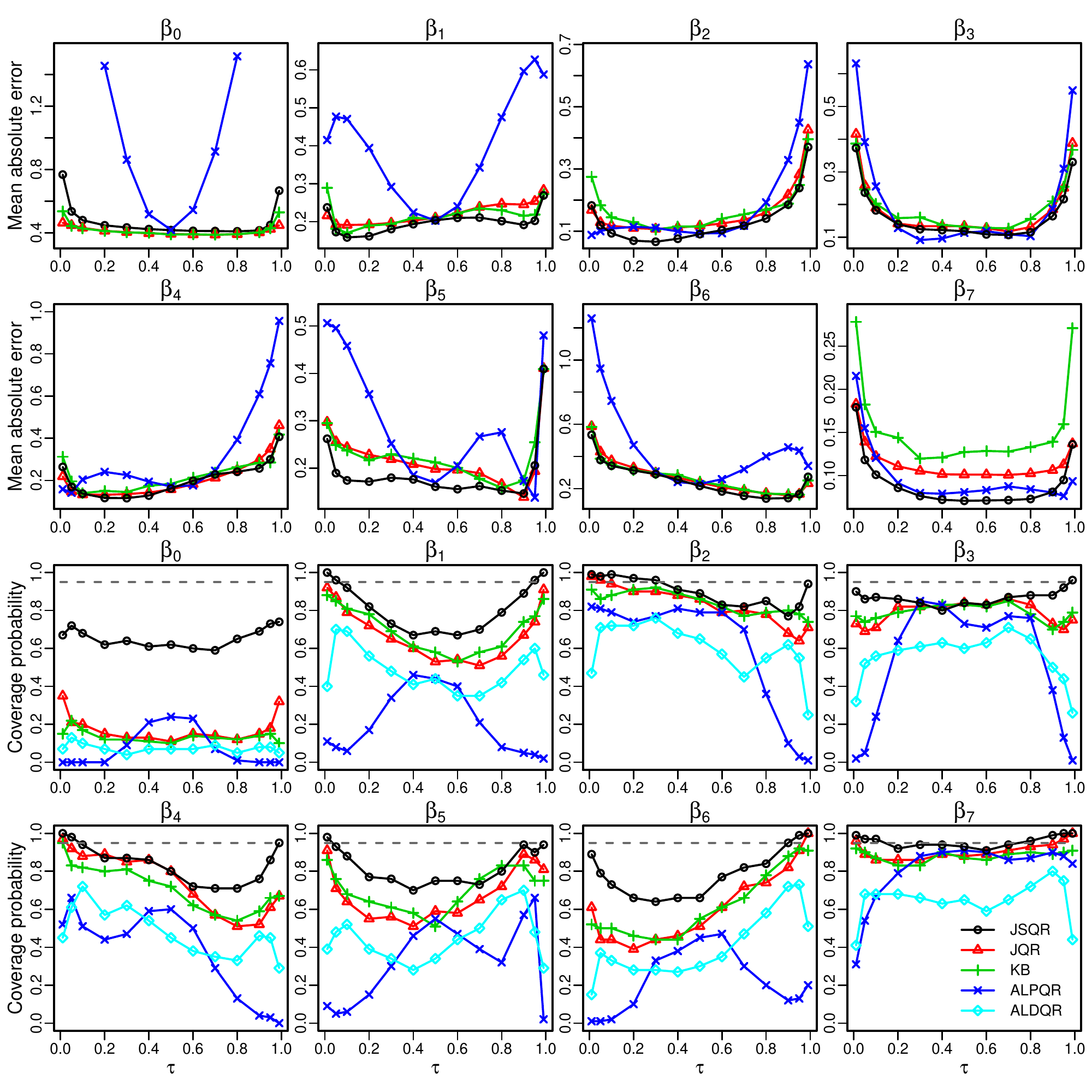}
\caption{\label{fig:ex2gpmaecp}Inference efficiency of different methods for Example 2 with true generating process being Gaussian process. The top two rows present the mean absolute errors of regression coefficients while the bottom two rows show the coverage probabilities of 95\% confidence (or credible) intervals of regression coefficients at $\tau\in\{0.01,0.05,0.1,\ldots,0.9,0.95,0.99\}$.} %\JSQR: proposed joint spatial quantile regression; \JQR: joint quantile regression by \cite{yang2017joint}; \KB: classical optimization-based method by \cite{koenker1978regression}; \ALPQR: spatial quantile regression using asymmetric Laplace error process by \cite{lum2012spatial}; \ALDQR: quantile regression using asymmetric Laplace error distribution by \cite{yu2001bayesian}.}
\end{figure}

\subsection{Conditional quantile function under the $t$ copula process}
\label{app:tprediction}
The conditional $t$ copula can be evaluated using conditional $t$ distribution. Let $s^*$ be a new location. Define $Z=\left(T^{-1}_\psi(U_1),\ldots,T_\psi^{-1}(U_n)\right)^\top$. Let $K$ be a $n\times n$ matrix with $K_{ij}=\rho(s_i,s_j;(\nu,\phi))$ and $K^*$ be a $n$-dimensional vector with $K^*_i=\rho(s^*,s_i;(\nu,\phi))$. Based on model specification \eqref{equ:tpmodel}, we have 
\begin{align*}
Z(s^*)\mid Z,\theta&\sim t_1\left(\psi + n, \mu(s^*), \frac{\psi + d}{\psi+n}\sigma^2(s^*)\right)
\end{align*}
where $d=Z^\top (\alpha K+(1-\alpha)I_n)^{-1}Z$, $\mu(s^*)=\alpha K^{*\top} (\alpha K+(1-\alpha)I_n)^{-1}Z$ and $\sigma^2(s^*)=1-\alpha^2 K^{*\top} (\alpha K+(1-\alpha)I_n)^{-1} K^*$. Therefore, the conditional $\tau^*$th quantile of response $Y^*$ given predictors $X^*$ at location $s^*$ is
\begin{align*}
     Q_{Y^*}(\tau^*\mid X^*,s^*,U_1,\ldots,U_n)&=\beta_0(\tau)+X^{*\top}\beta(\tau),~\tau=T_1\left(\mu(s^*)+\sqrt{\frac{\psi+d}{\psi+n}}\sigma(s^*)T^{-1}_{\psi + n}(\tau^*)\right)
\end{align*}

\subsection{Log-likelihood score with $W(s)$ and $\varphi$ as additional model parameters for $t$ copula process}
\label{app:loglik}
The \JSQR~model with $t$ copula process is given by
\begin{align*}
Y_i&=\beta_0(U_i)+X_i^\top\beta(U_i)\\
U_i&=T_\psi(Z(s_i)),~Z(s_i)=W(s_i)+\varepsilon(s_i)\\
W(s)&\sim\textsf{GP}(0,\alpha\rho(s,s';(\nu,\phi))/\varphi),~\varepsilon(s_i)\simiid \textsf{N}(0,(1-\alpha)/\varphi)\\
\varphi&\sim\Gam(\psi/2,\psi/2)
\end{align*}
Let $V_i=\Phi\left(\sqrt{\varphi/(1-\alpha)}\varepsilon(s_i)\right)$ and write the model conditioning on $W(s)$ and $\varphi$
\begin{align*}
Y_i&=\beta_{0}\left(h_{(W,\alpha,\varphi,\psi)}(s_i,V_i)\right)+X_i^\top\beta\left(h_{(W,\alpha,\varphi,\psi)}(s_i,V_i)\right),~V_i\simiid \textsf{Unif}(0,1),1\leqslant i\leqslant n
\end{align*}
where $h_{(w,a,\varphi,\psi)}(s,t)=T_\psi\left(w(s)+\sqrt{(1-\alpha)/\varphi}\Phi^{-1}(t)\right)$. Therefore, according to \eqref{equ:pdf}, the log-likelihood for $i$th observation with $W(s)$ and $\varphi$ as additional parameters is 
\begin{align}
-\log\left\{ \dot{\beta}_0(U_i)+X_i^\top\dot{\beta}(U_i)\right\}-\log \dot{h}_{(W,\alpha,\varphi,\psi)}(s_i,V_i)
\end{align}
where 
\begin{align*}
\dot{h}_{(W,\alpha,\varphi,\psi)}(s_i,V_i)=\frac{\sqrt{(1-\alpha)/\varphi}t_\psi(Z(s_i))}{\phi(\Phi^{-1}(V_i))}.
\end{align*}
The latent parameter $\varphi$ and the process realization $W(s)$ can be recovered according to their respective posteriors
\begin{align*}
    \varphi\mid Z(s), \theta,\psi&\sim\Gam\left(\frac{\psi+n}{2},\frac{\psi+Z(s)^\top(\alpha K+(1-\alpha)I_n)^{-1}Z(s)}{2}\right)\\
W(s)\mid Z(s),\theta,\varphi,\psi&\sim \textsf{N}\left(\frac{1}{1-\alpha}\left(\frac{1}{\alpha}K^{-1}+\frac{1}{1-\alpha}I_n\right)^{-1}Z(s), \frac{1}{\varphi}\left(\frac{1}{\alpha}K^{-1}+\frac{1}{1-\alpha}I_n\right)^{-1}\right)
\end{align*}
where $K$ is a $n\times n$ matrix with $K_{ij}=\rho(s_i,s_j;(\nu,\phi))$. We then have $V_i=\Phi(\sqrt{\varphi/(1-\alpha)}(Z(s_i)-W(s_i)))$.

\subsection{A comparison between \JSQR~and the model by \cite{reich2011bayesian}}
\label{subsec:svcillustration}
\begin{figure}[H]
\centering
\includegraphics[width=12cm]{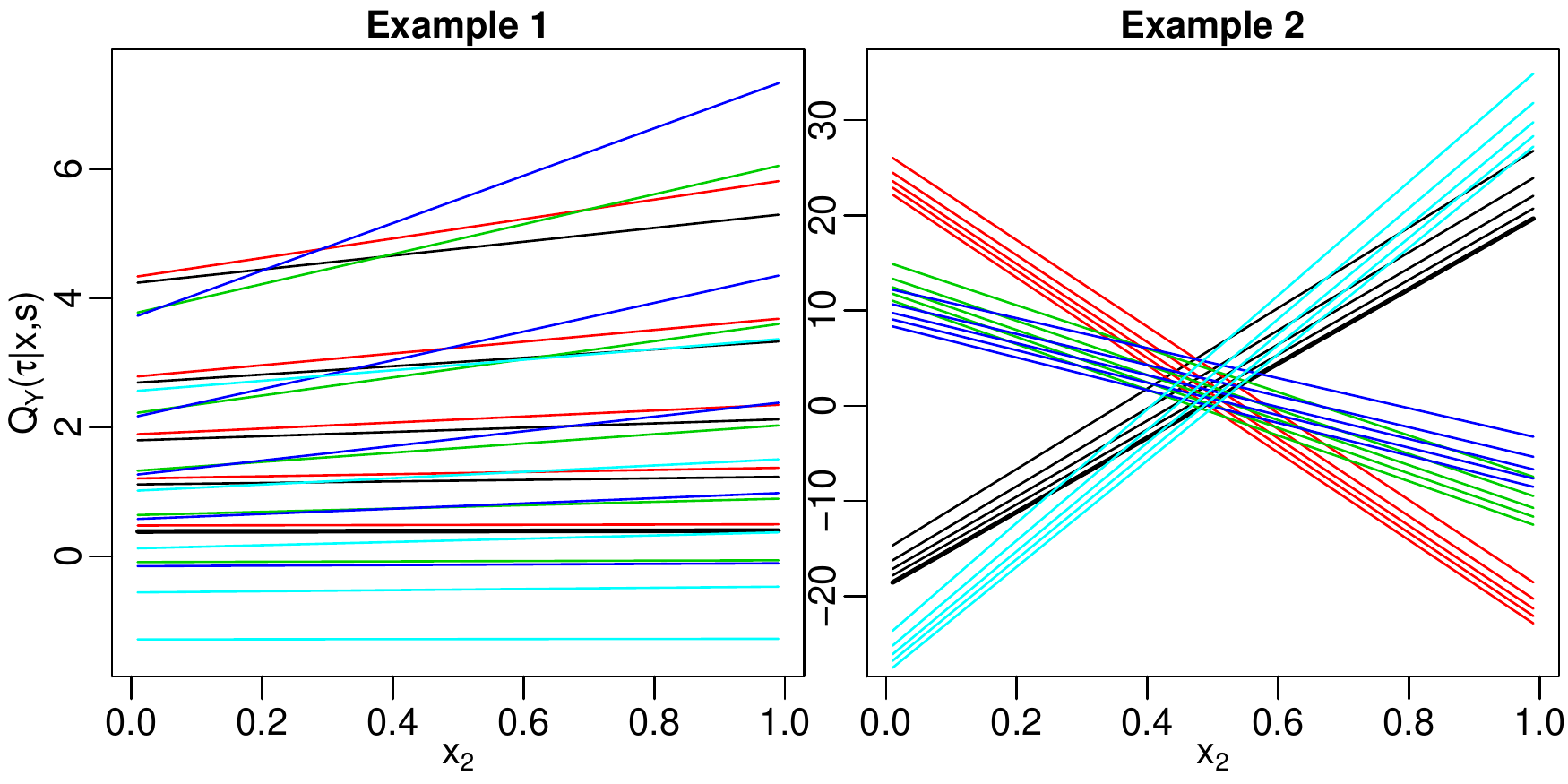}
\caption{\label{fig:quantilefun}Quantile functions $Q_Y(\tau\mid X,s)$ against $X_2$ at 5 randomly simulated locations with different colors at $\tau\in\{0.1,0.3,0.5,0.7,0.9\}$.}
\end{figure}

To thoroughly compare \JSQR~against \JQR~and the approximate spatial quantile regression (\ASQR) method proposed by \cite{reich2011bayesian}, we present two simulation studies with moderate and large spatial variations of regression coefficients respectively. In each simulation, we generated 80 replicates\footnote{\ASQR~required at least 79 observations at each location for estimating the covariance at $\tau=0.95$.} at each of 20 locations for each of the 100 synthetic datasets. For each replicate, we adopted the setup used in \cite{reich2011bayesian} to generate predictors, locations and quantiles.
\begin{align*}
 X_{i}\simiid\Un([0,1]^2),~s_{i}\simiid\Un([0,1]^2),~U_{i}=\Phi(Z(s_i)),~Z(s)\sim\textsf{GP}(0,\rho_{\text{SE}}(s,s';\sqrt{2}))
\end{align*}
The conditional quantile functions for simulation studies were
\begin{enumerate}[(i).]
\item $Q_{Y_{i}}(\tau\mid X_{i},s_{i})=2s_{i2}+(\tau+1)\Phi^{-1}(\tau)+5s_{i1}\tau^2X_{i2}$
\item $Q_{Y_{i}}(\tau\mid X_{i},s_{i})=2s_{i2}+(\tau+1)\Phi^{-1}(\tau)+5s_{i1}\tau^2X_{i2}+120(s_{i1}-0.5)(X_{i2}-0.5)$
\end{enumerate}
%\begin{flalign}
%\text{Example 1:}&& Q_{Y_{i}}(\tau\mid X_{i},s_{i})&=2s_{i2}+(\tau+1)\Phi^{-1}(\tau)+5s_{i1}\tau^2X_{i2}&\notag\\
%\text{Example 2:}&& Q_{Y_{i}}(\tau\mid X_{i},s_{i})&=2s_{i2}+(\tau+1)\Phi^{-1}(\tau)+5s_{i1}\tau^2X_{i2}+120(s_{i1}-0.5)(X_{i2}-0.5)&\notag\\
%&& X_{i}&\simiid\Un([0,1]^2),~s_{i}\simiid\Un([0,1]^2)&\notag\\
%&& U_{i}&=\Phi(Z(s_i)),~Z(s)\sim\textsf{GP}(0,\rho_{\text{SE}}(s,s';\sqrt{2}))&\notag
%\end{flalign}
The first example was taken from \cite{reich2011bayesian} where the true coefficients were $\beta_0(\tau,s)=2s_{i2}+(\tau+1)\Phi^{-1}(\tau),~\beta_1(\tau,s)=0$ and $\beta_2(\tau,s)=5s_{i1}\tau^2$. The quantile function is monotonically increasing in $X_{i2}$ with any fixed $(\tau,s)$. We added a term $120(s_{i1}-0.5)(X_{i2}-0.5)$ into the quantile function in the second example. As shown in Figure \ref{fig:quantilefun}, this added term greatly increased the variation of coefficients and quantile functions across locations. 

We specified a spatially varying version of \JQR~using $B$-splines (\JQRBS) of locations. Specifically, we adopted tensor product spline surfaces to approximate spatially varying coefficients
\begin{align*}
\hat{\beta}_k(\tau,s=(s_1,s_2))=\sum_{q=1}^d\sum_{r=1}^d\beta_{kqr}(\tau)\psi_q(s_1)\psi_r(s_2),~~k=0,\ldots,p
\end{align*}
where $\{\psi_r(\cdot): r=1,\ldots d\}$ are $B$-spline basis functions of order $d$. In our simulations, we augmented the original predictors (including intercept) by including their interactions with $\{\psi_q(s_1)\psi_r(s_2):q=1,\ldots,d,~r=1,\ldots,d\}$ resulting in $(d^2+1)(p+1)$ predictors in total. 

All settings for \JSQR, \JQR~were kept same as in Section \ref{sec:experiment}. We used $B$-splines with degree $d=2$ for $\JQRBS$. \ASQR~was implemented using the code available at the author's homepage\footnote{\url{https://www4.stat.ncsu.edu/~reich/code/}}. Instead of directly applying spectral decomposition on large correlation matrix as in Section \ref{sec:computation}, the computation of copula density of our model could be greatly simplified using Kronecker product. %(see Supplemental material \ref{app:svccomp}).
\JSQR~and \JQR~took 2.7 and 2 minutes respectively for analyzing one dataset on average while \JQRBS~and \ASQR~took 20 and 23.2 minutes. 

\begin{figure}[H]
\centering
\includegraphics[width=16cm]{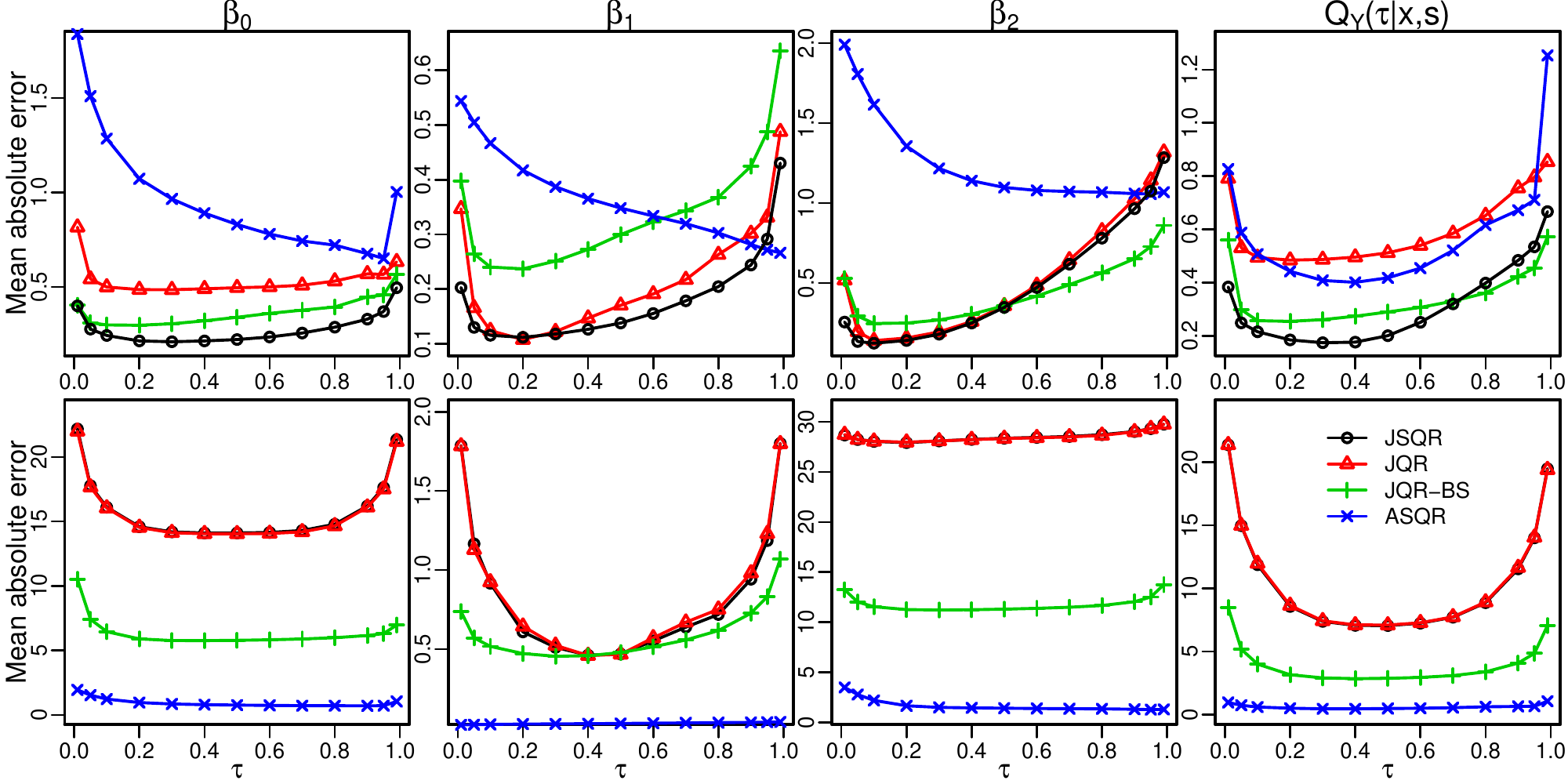}
\caption{\label{fig:svcsimu}Mean absolute errors of regression coefficients and quantile functions at $\tau\in\{0.01,0.05,0.1,\ldots,0.9,0.95,0.99\}$. The results for Example 1 and Example 2 are in the top and bottom rows respectively. %\JSQR: proposed joint spatial quantile regression; \JQR: joint quantile regression by \cite{yang2017joint}; \JQRBS: spatilly varying version of \JQR~using $B$-splines of locations; \ASQR: approximate spatial quantile regression method by \cite{reich2011bayesian}.}
}
\end{figure}

We compared different methods based on mean absolute errors of point estimates of regression coefficients and quantile functions. Namely, for each of 100 synthetic datasets, we calculated
\begin{align*}
&\frac{1}{m}\sum_{i=1}^{m}|\beta_k(\tau,s_i)-\hat{\beta}_k(\tau,s_i)|,~k=0,\ldots,p\\ 
&\frac{1}{mn}\sum_{i=1}^{m}\sum_{j=1}^{n}|Q_{Y_{ij}}(\tau\mid X_{ij},s_i)-\hat{Q}_{Y_{ij}}(\tau\mid X_{ij},s_i)|
\end{align*}
at quantile levels $\tau\in\{0.01,0.05,0.1,...,0.9,0.95,0.99\}$ and then computed the mean over all datasets. All estimates were given by posterior means. The simulation results are summarized in Figure \ref{fig:svcsimu}. In the first example, as shown in the left panel of Figure \ref{fig:quantilefun}, although the quantile functions from different locations got crossed, their slopes were similar. \JSQR~dominated all other competitors for estimating both coefficients and conditional quantiles. \JQR~provided the worst estimate for conditional quantiles while \ASQR~poorly estimated the coefficients. \JQRBS~indeed improved upon \JQR~according to MAE of conditional quantiles. Nevertheless, in the second example, the performances of \JSQR, \JQR~and \JQRBS~significantly dropped while the errors for \ASQR~were similar to those in the first example. The failure of \JSQR~and \JQR~in this example is not surprising due to the large spatial variation in regression coefficients. \JQRBS~significantly reduced errors by using splines of locations with degree 2. We expect that \JQRBS~has a better performance by using higher order splines. However, it also increases computational complexity. 

We did not include \JSQR~with augmented predictors using $B$-splines into comparison because its estimation accuracy was even worse than \JSQR~itself. The poor performance may be due to drastic overparameterization. Further investigation is needed to appropriately incorporate spatially varying coefficient into \JSQR~model. 

We note that the Example 2 is used to distinguish the behaviors of our method and \ASQR. In most of real data applications, we expect that the conditional distributions of the response given predictors at different locations have smaller spatial variation as in the Example 1 where our method is applicable.

\end{document}